\documentclass[%
]{article}

\usepackage{amssymb}
\usepackage{graphicx}
\usepackage{amsmath,amsfonts,amsthm,bm}
\usepackage{upgreek}
\usepackage{mathrsfs}
\usepackage{caption}
\usepackage{epstopdf}
\usepackage{tabularx}
\usepackage{hyperref}
\hypersetup{
    colorlinks=true,
    linkcolor=blue,
    filecolor=magenta,      
    urlcolor=cyan,
}
\usepackage{authblk}
\usepackage[
    letterpaper,
    top=2cm,
    bottom=2cm,
    left=2cm,
    right=2cm,
    textwidth=16cm,
    textheight=23cm
]{geometry}
\usepackage{hyperref}
\hypersetup{
    colorlinks=true,
    linkcolor=blue,
    citecolor=black,
    filecolor=magenta,      
    urlcolor=cyan,
}
\usepackage{mathtools}
\usepackage{amsthm}
\usepackage{natbib}
\usepackage{amsmath,etoolbox}
\usepackage{xcolor}
\usepackage{soul}
\usepackage{siunitx}
\usepackage{enumitem}
\usepackage[normalem]{ulem}

\usepackage[font=footnotesize,labelfont=bf]{caption}
\usepackage[labelformat=empty, position=top]{subcaption}
\usepackage[export]{adjustbox}
 
\theoremstyle{definition}
\newtheorem*{remark}{Remark}

\newcommand{\wtilde}[1]{\widetilde{#1}}

\newcommand{\jump}[1]{\left[\kern-0.15em\left[ #1 \right]\kern-0.15em\right]}  
\newcommand{\norm}[1]{\vert #1 \vert}

\newcommand{\I}{\mathbf{I}}

\newcommand{\Xv}{\mathbf{X}}
\newcommand{\xv}{\mathbf{x}}

\newcommand{\nv}{\mathbf{n}}
\newcommand{\Nv}{\mathbf{N}}

\newcommand{\Fb}{\mathbf{F}}
\newcommand{\FbT}{\Fb^{\text{T}}}

\newcommand{\Cb}{\bm{\mathbb{C}}}

\newcommand{\ev}{\mathbf{e}}
\newcommand{\uvb}{\mathbf{u}}

\newcommand{\bvb}{\mathbf{b}}
\newcommand{\hvb}{\mathbf{h}}
\newcommand{\mvb}{\mathbf{m}}
\newcommand{\Bvb}{\mathbf{B}}
\newcommand{\Avb}{\mathbf{A}}
\newcommand{\Hvb}{\mathbf{H}}
\newcommand{\Svb}{\mathbf{S}}
\newcommand{\sigvb}{\boldsymbol{\sigma}}

\newcommand{\muo}{\mu_{\texttt{0}}}

\newcommand{\Tmech}{\mathbf{T}}

\newcommand{\Diss}{\textit{D}}

\newcommand{\pt}{\texttt{p}}

\newcommand{\mecht}{\texttt{mech}}
\newcommand{\magt}{\texttt{mag}}

\newcommand{\vt}{\texttt{v}}

\newcommand{\foam}{\texttt{f}}
\newcommand{\sil}{\texttt{s}}
\newcommand{\pmt}{\texttt{pm}}

\newcommand{\Bvbr}{\boldsymbol{\mathcal{B}}^r}
\newcommand{\Hvbr}{\boldsymbol{\mathcal{H}}^r}
\newcommand{\Hvbrdot}{\dot{\boldsymbol{\mathcal{H}}}^r}

\newcommand{\p}{\partial}
\newcommand{\Vo}{{\mathcal{V}_0}}
\newcommand{\V}{{\mathcal{V}}}
\newcommand{\dVo}{{\partial \mathcal{V}_0}}
\newcommand{\dV}{{\partial \mathcal{V}}}

\newcommand{\Wmech}{\rho_\foam \Psi_\mecht}
\newcommand{\Wmag}{\rho_\foam \Psi_\magt}

\newcommand{\Wmechmat}{\rho_\pmt \Psi_\mecht^\pmt}

\newcommand{\db}{\delta b}
\newcommand{\dm}{\delta m}

\definecolor{mypink}{RGB}{180, 48, 122}

\begin{document}

\title{Experiments and modeling of mechanically-soft, hard magnetorheological foams with potential applications in haptic sensing}

\author[1]{Zehui Lin}

\author[2]{Zahra Hooshmand-Ahoor}

\author[1]{Laurence Bodelot\thanks{Email: \texttt{laurence.bodelot@polytechnique.edu}}}

\author[1]{Kostas Danas\thanks{Email: \texttt{konstantinos.danas@polytechnique.edu}}}

\affil[1]{LMS, CNRS, \'Ecole Polytechnique, Institut Polytechnique de Paris, Palaiseau, 91128, France}

\affil[2]{Department of Engineering, University of Cambridge, Trumpington Street, Cambridge, UK}

\maketitle 

\begin{abstract}
This study proposes a family of novel mechanically-soft and magnetically-hard magnetorheological foams that, upon deformation, lead to robust and measurable magnetic flux changes in their surroundings. This allows to infer qualitatively and even quantitatively the imposed deformation and, eventually from that, an estimation of the stiffness and average stress on the sample even in complex loading scenarios involving combinations of uniform or nonuniform compression/tension with superposed shearing in different directions. The work provides a complete experimental, theoretical and numerical framework on finite strain, compressible magneto-elasticity, thereby allowing to measure and predict coupled magneto-mechanical properties of such materials with different particle volume fractions and then use it to estimate and design potential haptic sensing devices.


\end{abstract}


\section{Introduction}

Magnetorheological elastomer (MRE) foams have gained increased interest in the last few years,  and are a special class of porous MREs that exhibit a highly compressible response. Contrary to more conventional incompressible MREs, MRE foams may prove more versatile in a number of cases involving actuation but also sensing, which is the main contribution of the present work. 

Specifically, MRE foams, owing to their lightweight nature \citep{DAuria2016} can exhibit improved deformation and magnetostriction performance, and larger relative magnetorheological (MR) effect than pure MREs \citep{Ju2013}. Such foams are mainly fabricated through two methods \citep{Muhazeli2021}. The first method, called \textit{ex-situ}, impregnates preformed foam matrices with MRE composites \citep{Ge2015,Li2019}. But most studies use the \textit{in-situ} way that incorporates magnetic particles within the base matrix, and then generates pores via a foaming agent before curing \citep{Plachy2018,Norhaniza2020}. In-situ MRE foams have been explored through the use of polyurethane foams loaded with carbonyl iron powder, referred to as  $s$-MRE foams. Researchers have explored their morphological and rheological properties \citep{Schumann2015, Muhazeli2019}, magneto-mechanical response \citep{Diguet2021}, anisotropic mechanical enhancements \citep{Sorrentino2011, Gong2013}, and shock energy dissipation \citep{Ju2013}. In most of those studies, the effort focused on increasing the magnetorheological effect \citep{Diguet2022} and relevant magnetostrictive capability, on noise reduction \citep{Muhazeli2019, Muhazeli2020} or on  adaptive cushioning \citep{Choi2024}, i.e., using the foams in the actuating regime. Yet, in \cite{Diguet2022}, it becomes clear that the change in magnetic permeability with the application of mechanical loads is a potential mechanism allowing to use such materials for sensing instead of actuation. Nevertheless, a major difficulty in using  directly $s$-MRE foams for sensing is the necessity of applying an external magnetic field (usually large) to magnetize such samples and operate them. 

Alternatively, the use of $h$-MRE materials could fill this gap since they display magnetic field changes upon deformation without the need of an externally applied magnetic field (except during premagnetization). Recently, however, it was shown numerically and theoretically \citep{mukherjee2021,mukherjee2022,DanasReis2024} as well as experimentally \citep{Yan2023} that $h$-MREs exhibit a stretch-independent magnetization response upon deformation, thus making them \textit{a priori} less interesting candidates for the inverse sensing process. Moreover, as stated in \cite{zhao2019} and \cite{Diguet2022}, upon full pre-magnetization, $h$-MREs exhibit a relative permeability close to unity, thus magnetic changes are expected to be small even at relatively finite deformations. 

As a first effort towards that direction, we have proposed in \cite{Lin2025} a new class of $h$-MRE foams comprising ellipsoidal voids whose largest axis lies along the foaming direction. In that first study, we have only studied the magnetization saturation response of the foams showing that it is a linear function of the overall particle fraction and the magnetization saturation of the particles. Also, we have shown that the anisotropic shape of the voids does not affect the magnetization response of the $h$-MRE foams, i.e. they exhibit isotropic magnetic properties in the absence of mechanical loads. Nonetheless, void anisotropy is expected to affect their mechanical response and for that reason no further effort was done along this direction, since it would have required a very large number of experiments as well as a rather complex modeling approach.

Instead, this work proposes a novel and complete theoretical, numerical and experimental framework for the magneto-mechanical response of \emph{isotropic} $h$-MRE foams. The fabrication process of the present $h$-MRE foams builds upon the first study of \cite{Lin2025} but is slightly modified to obtain both isotropic mechanical and magnetic responses (i.e. the resulting voids have rather polygonal and spherical shapes). This, in turn, allows to simplify substantially the rather complex compressible magneto-mechanical response of these materials. In particular, we provide a combined experimental protocol coupled with analytical solution of the boundary value problem corresponding to the experimental setup and a fully explicit homogenization-based model with only a very small number of material parameters requiring calibration. Furthermore, we show that these permanently magnetized foams provide versatile sensing capabilities without the need for an externally-applied magnetic field by taking advantage of the pore closure/opening and the resulting effective change in apparent magnetization saturation, but also of the specimen shape changes involved in such large deformation processes. 

The present work is organized as follows. In Section~\ref{sec:Kinematics and magnetostatics}, we first present some preliminary definitions for the kinematics and magnetostatics that are helpful to proceed with the discussion of the fabrication and experimental study of the $h$-MRE foams in Section~\ref{sec:Fabrication and experiments}. Subsequently, in Section~\ref{sec:Thermodynamics, constitutive modeling and calibration}, we present an analytical, thermodynamically-consistent, dissipative magneto-mechanical model for compressible $h$-MRE foams, along with some simplifications of it. In Section~\ref{sec:Model calibration and predictive capabilities by comparison with the experiments}, we calibrate in a fully analytical manner (i.e., without the use of numerical simulations) a very small number of constants related to the magneto-dilatational coupling, and compare the model predictions with the available experimental data for various particle volume fractions and resulting porosities. Section~\ref{sec:Numerical study: multi-loading sensing capabilities of hMRE foam} provides a numerical study of pragmatic boundary value problems allowing to probe the versatility of the present $h$-MRE foam as a potential haptic/tactile sensor. We close the study with concluding remarks in Section~\ref{sec:Concluding remarks}.

\section{Kinematics, magnetostatics and microstructural definitions}
\label{sec:Kinematics and magnetostatics}

We present first the basic kinematics and magnetostatics by considering a homogeneous magneto-active solid to simplify the discussion and maintain generality. The subsequent constitutive description of the foam uses a combination of known homogenization results at both scales, i.e., that of the microscopic elastomer matrix-particle scale and that of the mesoscopic $h$-MRE matrix and voids scale. In addition, we neglect in this study any dissipative phenomena arising from the viscoelasticity of the polymer phase, as well as inertia terms. 

The boundary of the solid is assumed to be smooth and is designated by $\dVo$ ($\dV$), while $\Nv$ ($\nv$) denotes the unit normal on $\dVo$ ($\dV$) in the reference (current) configuration. In the present formulation, we consider as primary variables the mechanical displacement field $\uvb(\Xv)$, which relates the current, $\xv$, and reference, $\Xv$, position vectors via $\xv = \Xv + \uvb(\Xv)$, and the magnetic vector potential field $\Avb(\Xv)$ defined in the reference configuration. This allows to write the deformation gradient $\Fb$ and Lagrangian magnetic flux $\Bvb$ as
\begin{equation}
\Fb = \I + \mathrm{Grad}\,\uvb(\Xv), \qquad \Bvb(\Xv)=\mathrm{Curl}\, \Avb(\Xv).
\label{eq:F-B_definitions}
\end{equation}
Here, $\I$ is the second-order identity tensor, and both $\mathrm{Grad}$ and $\mathrm{Curl}$ operators are defined with respect to the reference position $\Xv$. For later use in presenting the various results, we also introduce the engineering axial strain $\varepsilon_{ii}=F_{ii}-1$ (no sum on $i$) and shear strains $\gamma_{12}=F_{12}$ and $\gamma_{32}=F_{32}$ and we define the right Cauchy-Green deformation tensor $\Cb$ and the volume Jacobian $J$ as
\begin{equation}
\Cb=\FbT \Fb, \qquad J = \det \Fb = \sqrt{\det\Cb}  > 0.
\label{eq:C-J_detF_def}
\end{equation}
The inequality in this last expression serves to impose the impenetrability condition. The importance of this last definition in the present work is paramount since we deal with compressible foams. The above definitions satisfy automatically the compatibility conditions, the absence of magnetic monopole condition as well as jump conditions across perfect interfaces (or boundaries), i.e.,
\begin{align}
\textrm{Curl}\, \Fb=\mathbf{0}, \qquad \jump{\Fb}\times \Nv=\mathbf{0}, \qquad \mathrm{Div}\,\Bvb=0,\qquad \jump{\Bvb}\cdot \Nv=0.
\end{align}

We introduce next the first Piola or engineering stress tensor $\Svb$ and the magnetic field strength (or H-field) $\Hvb$ as conjugate fields to $\Fb$ and $\Bvb$, respectively. Conservation of linear and angular momentum (in absence of mechanical body forces and inertial effects) as well as Amp\`ere's law (in absence of electric currents or charges) lead to the pointwise differential equations and jump conditions across perfect interfaces (or boundaries), such that
\begin{equation}
	\mathrm{Div} \, \Svb = \bm{0}, \qquad \jump{\Svb}\Nv-\Tmech = \bm{0}, \qquad \mathrm{Curl}\,\Hvb = \mathbf{0}, \qquad \jump{\Hvb}\times \Nv = \mathbf{0}.
	\label{eq:Ampere_law_pde_reference}
\end{equation}
Here, $\Tmech$ denotes the mechanical traction in the reference configuration. 

Use of standard push-forward transformations allows to write the stress and magnetic fields in the current configuration, i.e., their Eulerian counterparts read \citep{ogden1997,dorfmann2004nonlinear,kankanala2004finitely}
\begin{equation}
\sigvb = \dfrac{1}{J} \Svb \FbT, \qquad \bvb = \dfrac{1}{J} \Fb \Bvb, \qquad \hvb = \Fb^{-\text{T}}\Hvb. \label{eq:transforms_shb}
\end{equation} 
For later use in presenting the results, one may also define the current magnetization vector $\mvb$ via the constitutive relation
\begin{equation}
	\bvb=\mu_0 (\hvb+\mvb) \qquad \mathrm{or} \qquad \mvb=\dfrac{1}{\mu_0}\bvb-\hvb, 
	\label{eq:magnetization_definition}
\end{equation}
where $\mu_0$ is the magnetic permeability of vacuum, air or non-magnetic solids. 

This last expression serves as a \emph{non-unique definition} of the magnetization vector in the current volume $\V$, which however is not defined on its boundary $\dV$. In fact, it was recently shown theoretically and experimentally \citep{mukherjee2021,mukherjee2022,Yan2023,Danas2024,DanasReis2024} that $\mvb$ may be directly related to an internal state variable in the general case of dissipative magnetic solids, as is the case here. In the present case of compressible magnetorheological materials, we will show even more clearly that magnetization is a complex function of the internal magnetic variable (to be defined subsequently), the deformation gradient and the Jacobian $J$ defined in equation \eqref{eq:C-J_detF_def}.

The fundamental quantities introduced in this section will be used in the following to report the experimental results and construct the magneto-mechanical theory.

\section{Experiments}
\label{sec:Fabrication and experiments}

\subsection{Sample preparation} 
\label{subsec:Material selection and fabrication protocol}

The material used in this study is obtained following most of the steps described in the recent work of \cite{Lin2025}, with the main difference lying in the fabrication of isotropic samples contrary to anisotropic ones in that earlier work. Specifically, we fabricate the $h$-MRE foam using as a base the commercial Soma Foama 25 (Smooth-On), which is a soft two-component (Part A and B) flexible platinum silicone foam, mixed with NdFeB hard magnetic particles in form of powder (MQFP-16-7, Neo Magnequench) with a mean diameter (MV) of $5~\si{\micro\meter}$.

\begin{figure}[h!]
  \centering
  \includegraphics[width=0.9\textwidth]{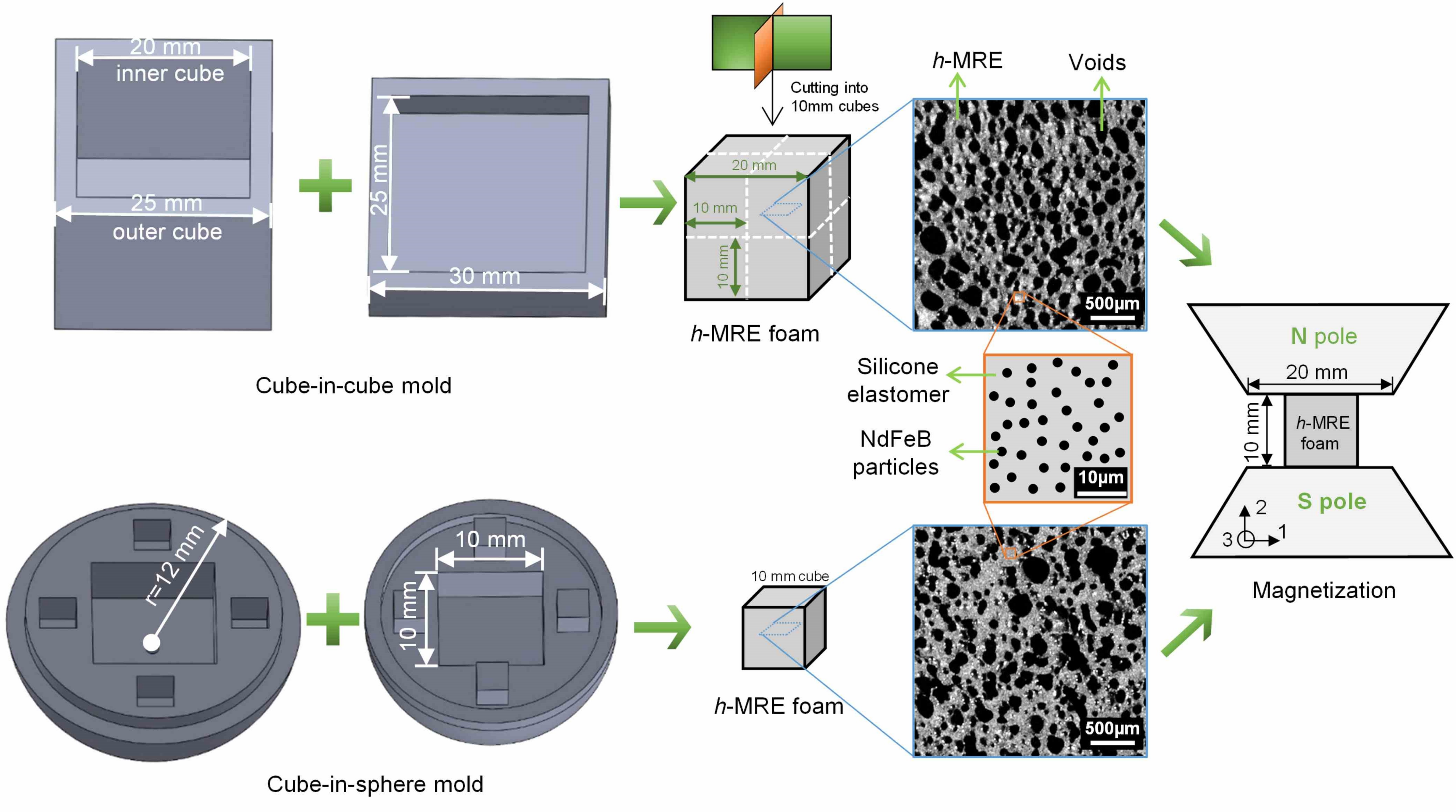}
  \caption{Geometric dimensions of (top row) cube-in-cube and (bottom row) cube-in-sphere molds. From the cube-in-cube we obtain 8 specimens of volume $10\times 10\times 10~\si{\milli \meter}^3$. From the cube-in-sphere mold we obtain one specimen of the same dimensions. Optical microscope images of the as fabricated $h$-MRE foams show a polydisperse void distribution with sizes ranging between $50{-}200~\si{\micro \meter}$. A sketch of the silicone matrix with the magnetic particles is shown also for completeness. The $h$-MRE foam samples are subsequently permanently magnetized using a two-coil electromagnet. }
  \label{fig:molds}
\end{figure}
\begin{figure}[h!]
  \centering
  \begin{subfigure}[t]{0.02\textwidth}
    {\small a)}
  \end{subfigure}
  \begin{subfigure}[t]{0.43\textwidth}
    \includegraphics[width=\linewidth, valign=t]{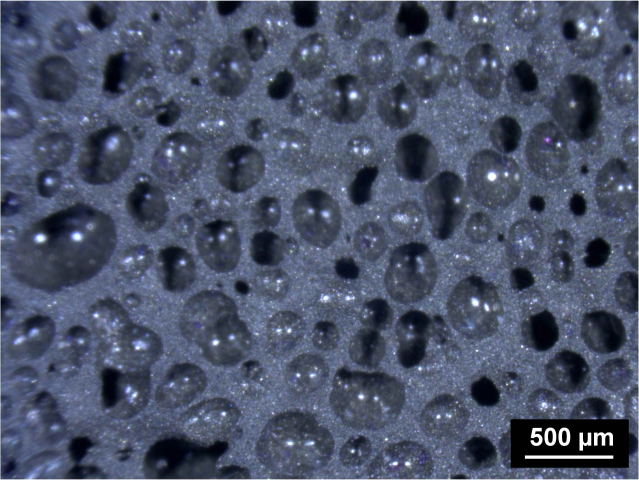}
  \end{subfigure}\hfill
  \begin{subfigure}[t]{0.02\textwidth}
    {\small b)}
  \end{subfigure}
  \begin{subfigure}[t]{0.43\textwidth}
    \includegraphics[width=\linewidth, valign=t]{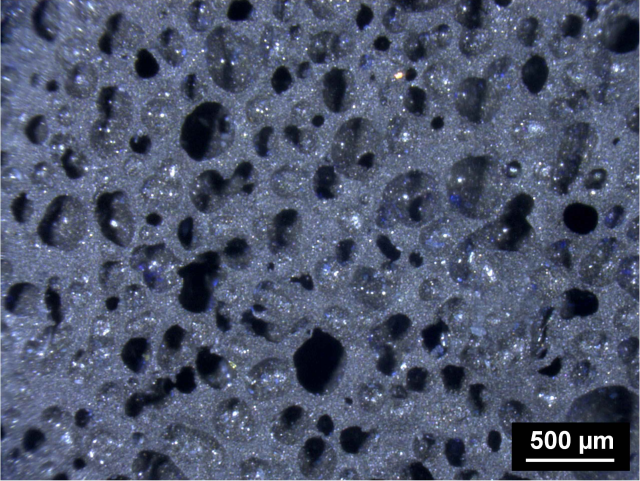}
  \end{subfigure}  
  \\[2ex]
  \begin{subfigure}[t]{0.02\textwidth}
    {\small c)}
  \end{subfigure}
  \begin{subfigure}[t]{0.43\textwidth}
    \includegraphics[width=\linewidth, valign=t]{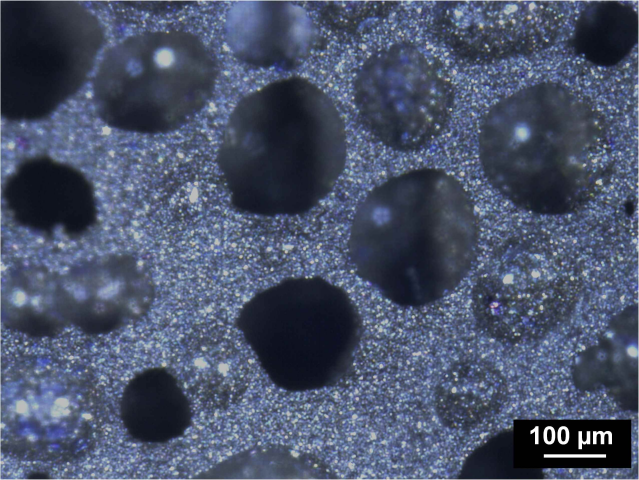}
  \end{subfigure}\hfill
  \begin{subfigure}[t]{0.02\textwidth}
    {\small d)}
  \end{subfigure}
  \begin{subfigure}[t]{0.43\textwidth}
    \includegraphics[width=\linewidth, valign=t]{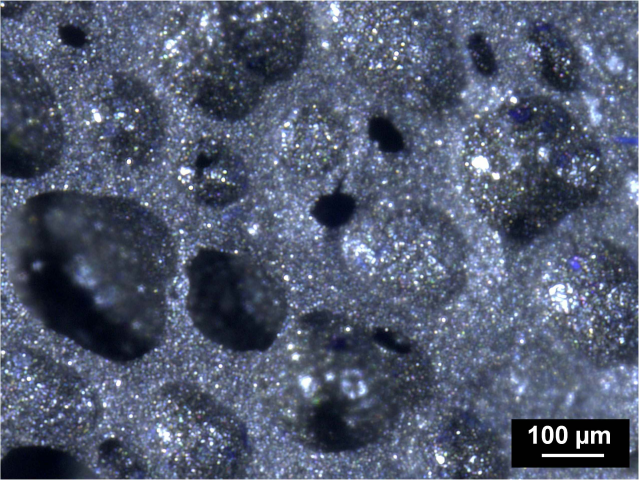}
  \end{subfigure}\hfill
  \caption{High resolution optical micrographs of the $h$-MRE foam microstructure with closed porosity and $c_\pt=0.12$ (see Table~\ref{tab:microstructural_params} for more details on the microstructural parameters). (a,c) horizontal and (b,d) vertical cross-sections of the samples. (c,d) closeup zooms revealing the NdFeB particles as white spots due to light reflections.}
  \label{fig:micrographs}
\end{figure}
For the fabrication of the $h$-MRE foam, we first disperse the particles in Part B (which is the most viscous component) along with Silicon Thinner fluid (Smooth-On) to lower the overall viscosity and promote a homogeneous dispersion of the particles. The compound is thoroughly hand-mixed for 15~s. Then, we add Part A, and mix again for 15~s before pouring the mixture in a mold previously sprayed with a release agent (Ease Release 200, Smooth-On) in order to facilitate the removal of the sample after curing. Contrary to the previous study of \cite{Lin2025}, where free foaming in open molds led to elongated ellipsoidal voids along the foaming direction, the goal in this study is to obtain composite foams that are as isotropic as possible. To achieve this, we have used two closed molds for constrained foaming consisting of a cube-in-cube and a cube-in-sphere geometries as shown in Fig.~\ref{fig:molds}. After pouring the compound in the molds and letting it rest for 5~min, we rotate the first mold by hand and the second using a salad mixer at low speed. This rotation is done for 15 min and allows for bubble growth in all directions. Large rotation speeds lead to a central large bubble cavitation in the center of the specimen and thus should be avoided. Finally, the samples are left at room temperature for 1~h to solidify before being unmolded. We note that the use of two molds is done for comparison purposes but also to provide two alternative methods that ultimately lead to very similar microstructures and results, as we discuss below.  

Fig.~\ref{fig:molds} also shows optical microscope images of samples obtained with the two different molds. The voids exhibit convex polygonal shapes with sizes ranging between $50{-}200~\si{\micro \meter}$. For additional precision, Fig.~\ref{fig:micrographs} shows horizontal and vertical cuts performed in two samples with total particle volume fraction $12vol\%$ indicating a closed porosity microstructure with voids uniformly distributed in the $h$-MRE matrix phase. The high-magnification images in Fig.~\ref{fig:micrographs}c,d allow to distinguish a well-distributed particle phase (see white spots due to light reflection) that lies at substantially smaller length scale than that of the voids. The expected ``isotropic'' mechanical and magneto-mechanical response of those two sets of samples is discussed in detail in the next Section~\ref{subsec:Experimental results}.

\subsection{Microstructural considerations}
\label{subsec:Microstructural considerations}

Guided by those experimental micrographs, we consider that the $h$-MRE foam occupies a volume $\Vo$ ($\V$) in its reference (current) configuration. In addition, it may be regarded as a three-phase material made up of an incompressible $h$-MRE matrix phase and voids at a mesoscopic scale (i.e., void size of about $50-200~\si{\micro\meter}$ as shown in Fig.~\ref{fig:micrographs}a,b). The $h$-MRE matrix phase itself consists of an elastomer silicone matrix and hard magnetic particles at a lower microscopic scale (i.e., particle size of average diameter $5~\si{\micro \meter}$ as shown in Fig.~\ref{fig:micrographs}c,d). Assuming sufficient separation of the aforementioned scales, the composite foam is then modeled as a homogeneous magneto-active foam at the macroscopic ($\si{\milli \meter}$) scale, as discussed in Section~\ref{sec:Thermodynamics, constitutive modeling and calibration}.

However, before we proceed with specific constitutive laws, it is necessary to provide some preliminary definitions for the relevant volume fractions of the phases and their estimation from the experimental samples. As has been shown experimentally and theoretically in \cite{Lin2025}, the main parameter affecting the magnetic response of an $h$-MRE foam is the total volume fraction of particles therein. For this reason, we define next the reference total volume of the foam as $\V_0^\foam=\V_0^\vt+\V_0^\pt+\V_0^\sil$, with $\V_0^\vt$ being the reference volume of the voids, $\V_0^\pt$ of the particles, and $\V_0^\sil$ of the pure silicone. We introduce next the reference volume fraction $c_\vt$ of the voids, $c_\pmt$ of the particles in the $h$-MRE matrix phase (comprising the particles and the silicon but not the voids), and $c_\pt$ of the particles in the entire foam. Using those definitions, one obtains the following universal relations
\begin{equation}
	c_\vt= \dfrac{\V_0^\vt}{\V_0^\foam}, \qquad c_\pmt = \dfrac{\V_0^\pt}{\V_0^\pt+\V_0^\sil}, \qquad c_\pt=\dfrac{\V_0^\pt}{\V_0^\foam}= \dfrac{\V_0^\pt}{\V_0^\pt+\V_0^\sil}\, \dfrac{\V_0^\pt+\V_0^\sil}{\V_0^\foam}=c_\pmt \,(1-c_\vt).
	\label{eq:cv_cpm_cp_defs}
\end{equation}
This last set of equations directly implies that, unlike the more common case of incompressible $h$-MRE materials, in $h$-MRE foams, the particle volume fraction is a linear function of the porosity, which, in turn, evolves with mechanical deformation. In simple terms, as the volume of the foam decreases (increases) due to mechanical loads, the overall volume fraction of the particles in the foam increases (decreases). It is fairly straightforward to show \citep{Hooshmand2022} that the current porosity and particle volume fraction, denoted as $\wtilde{c}_\vt$ and $\wtilde{c}_\pt$, respectively, in the deformed configuration can be expressed in terms of the reference $c_\vt$, $c_\pt$ and $c_\pmt$ as
\begin{equation}
\wtilde{c}_\vt=1-J^{-1}(1-c_\vt), \qquad \wtilde{c}_\pmt=c_\pmt, \qquad \wtilde{c}_\pt=\wtilde{c}_\pmt (1-\wtilde{c}_\vt)=J^{-1} c_\pmt (1-c_\vt)=J^{-1} c_\pt.
\label{eq:cv_cp_cpm_current_def}
\end{equation}
In these last expressions, we have taken into account the incompressibility of the $h$-MRE matrix phase, which implies that $c_\pmt$ does not change with the deformation and thus with $J$. The subsequent constitutive laws need to properly take into account those relations wherever needed.

\begin{table}[h]
\centering
\caption{Microstructural measurements}
\begin{tabular}{c c c c} \hline
$c_\pt$ & $c_\vt$ & $c_\pmt$ \\ \hline
0.04 & 0.51 & 0.08\\
0.08 & 0.49 & 0.15 \\ 
0.12 & 0.45 & 0.21 \\ \hline 
\end{tabular}
\label{tab:microstructural_params}
\end{table}
Using the above definitions and following the procedure detailed in \cite{Lin2025}, we fabricate three sets of samples with different initial particle volume fraction $c_\pmt$ in the $h$-MRE matrix phase leading to different $c_\vt$ and $c_\pt$ in the foam, as reported in Table~\ref{tab:microstructural_params}. The porosity $c_\vt$ is estimated from the density of the samples by simply measuring their mass and volume. For these estimations, we have used a density of $1,021~\si{\kilogram / \meter^3}$ for the pure silicone matrix phase as obtained experimentally by weight measurements of known volumes of fluid components, while the density of the particles is $7,610~\si{\kilogram / \meter^3}$ as provided by the manufacturer. It is noted that the values reported in Table~\ref{tab:microstructural_params} are averages from four different samples under each set, with porosity exhibiting a variation of $\pm 0.01$ in all cases considered here. This leads to a $\pm0.005$ variation in the overall particle volume fraction $c_\pt$ in the foam.

\subsection{Sample testing}
\label{subsec:Experimental results}

In this section, we discuss the experimental testing of magnetized $h$-MRE foam samples under compressive loads. First, we magnetize permanently the $h$-MRE foam samples under a strong external magnetic field with a maximum strength of $3.1~\si{\tesla}$ applied at a rate of 0.04~T/s using a two-coil electromagnet (Bouhnik) equipped with iron poles of $20~\si{\milli \meter}$ diameter separated by an air gap of $10~\si{\milli \meter}$  (see magnetization sketch in Fig.\ref{fig:molds}). The surrounding magnetic field is uniform ($<$ 1\% deviation) allowing to reach magnetization saturation everywhere in the sample and thus uniform magnetization upon removal of the applied magnetic field.

Subsequently, the samples are taken out of the magnet and are subjected to two types of compression tests. The compression device generally consists of a fixed bottom plate, and a top plate attached to a linear displacement motor (DRLM42G-04B2M-K, Oriental motors) operated at a rate of $0.02~\si{\milli \meter / \second}$ . To eliminate the effect of the compliance of the system, the displacement applied to the sample is measured by tracking markers located on the plates (or piston) directly in contact with the sample thanks to a camera (GO-5100M-USB, Jai) equipped with a telecentric lens (FXL-0305-VT-165, Seiwa optical). The force exerted on the sample is measured with a force sensor (LCAE-10KG, Omega Engineering). Finally, a 3-axis Hall sensor (F3A-03HSs02C-A.1T2K5M, Senis), placed underneath the bottom face of the sample at its center (with a vertical offset of $0.5~\si{\milli \meter}$), measures all three components of the magnetic flux at that point. 

\begin{figure}[h!]
  \centering
  \begin{subfigure}[t]{0.02\textwidth}
    {\small a)}
  \end{subfigure}
  \begin{subfigure}[t]{0.9\textwidth}
    \includegraphics[width=\linewidth, valign=t]{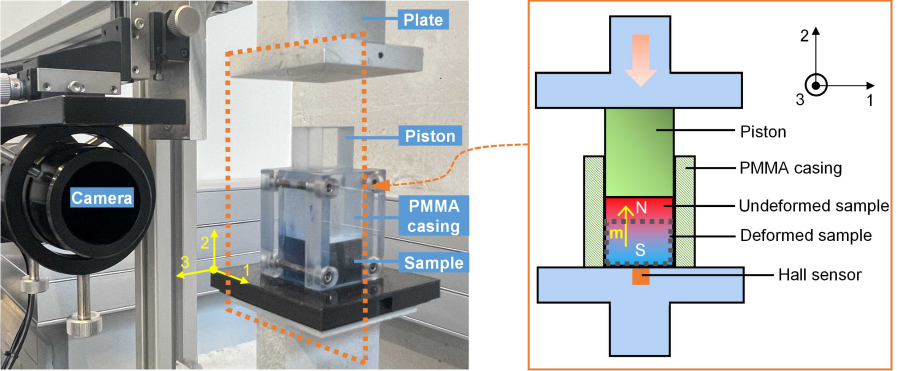}
  \end{subfigure}\\[2ex]
  \begin{subfigure}[t]{0.02\textwidth}
    {\small b)}
  \end{subfigure}
  \begin{subfigure}[t]{0.9\textwidth}
    \includegraphics[width=\linewidth, valign=t]{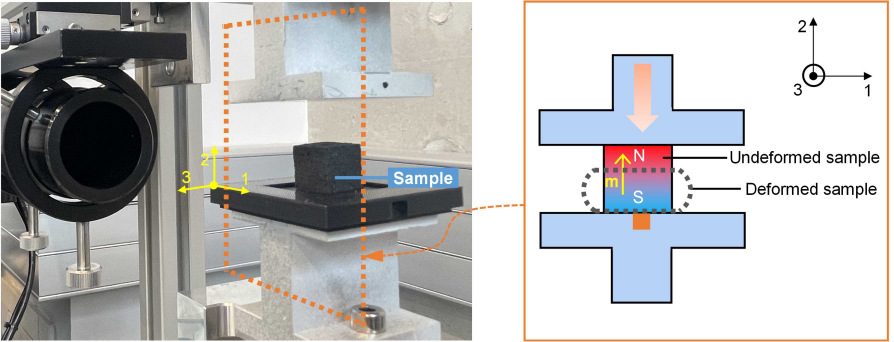}
  \end{subfigure}\hfill
  \caption{Two experimental tests are considered. A camera tracks markers to measure the sample deformation, a force sensor (not shown) attached to the top plate measures the force applied to the sample, and a 3-axis Hall sensor measures the magnetic flux $0.5~\si{\milli \meter}$ below the center of the bottom side of the sample (see sketches in orange color). All measurements are synchronized in time. (a) Oedometric compression along the magnetization direction~$2$. A PMMA casing with a square cross-section of $10\times10~\si{\milli \meter}^{2}$ constrains the lateral displacements of the sample. (b)~Uniaxial compression along the magnetization direction $2$ where the lateral sides are traction-free.}
  \label{fig:exp_setup}
\end{figure}

The first test is an oedometric compression along the magnetization direction. In this case, the lateral normal displacements of the sample are constrained by use of a poly(methyl methacrylate) (PMMA) casing while the vertical displacement is applied by a piston pushed within the casing by the top plate of the compression device (see Fig.~\ref{fig:exp_setup}a), thus leading to an overall uniaxial strain state. The data obtained from the oedometric experiment will be used primarily to calibrate the magneto-mechanical model proposed in Section~\ref{sec:Thermodynamics, constitutive modeling and calibration}. The main advantage of this test is the fairly simple resulting stress and strain fields (under certain assumptions such as frictionless contact of the sample with the side walls), thus allowing for an analytical treatment of both the mechanical and magneto-mechanical responses as discussed later in Section~\ref{sec:Model calibration and predictive capabilities by comparison with the experiments}. 

The second test, shown in Fig.~\ref{fig:exp_setup}b, is a uniaxial compression along the magnetization direction $2$, where the top and bottom faces of the sample are in contact with both plates of the compression device and the lateral sides are traction-free. This configuration leads to nonuniform compressive stress and strain states, with the lateral sides of the sample bulging substantially. The experimental data from this test will be used in Section~\ref{sec:Model calibration and predictive capabilities by comparison with the experiments} to probe the predictive capability of the proposed model. 

In the following, we discuss briefly representative experimental data obtained for the set of samples comprising the largest particle volume fraction, i.e., $c_\pt=0.12$ (see Table~\ref{tab:microstructural_params}). Specifically, we show the absolute value of the engineering (or first Piola) compressive stress, $|S_{22}|$, the current magnetic flux $b_2$ and the variation of the magnetic flux $\db_2$ along the magnetization direction as a function of the applied absolute engineering compressive strain, $|\varepsilon_{22}|$. This average strain is defined simply as the applied displacement divided by the initial height of the sample. For later use in the proposed magneto-mechanical model and its calibration, we also show stress-strain data for unmagnetized samples. We use a total of four samples with $c_\pt=0.12$, two from each mold described earlier, to extract the magneto-mechanical response. In addition, for the extraction of the stress response in the unmagnetized samples, we consider compression tests along all three principal directions of the cube, allowing this way to probe the mechanical isotropy of the response. 

\begin{figure}[h!]
  \centering
  \begin{subfigure}[t]{0.02\textwidth}
    {\small a)}
  \end{subfigure}
  \begin{subfigure}[t]{0.3\textwidth}
    \includegraphics[width=\linewidth, valign=t]{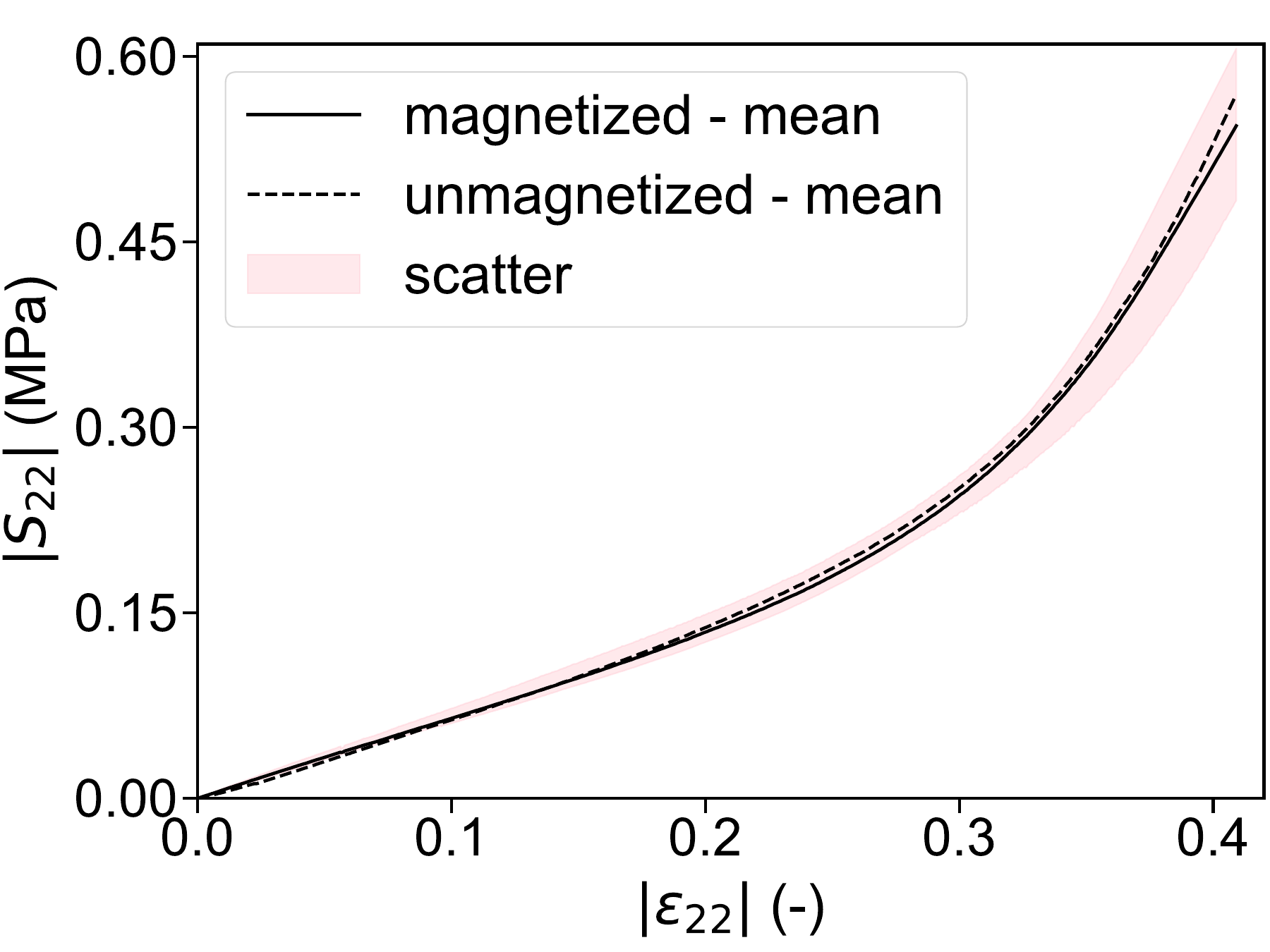}
  \end{subfigure}\hfill
  \begin{subfigure}[t]{0.02\textwidth}
    {\small b)}
  \end{subfigure}
  \begin{subfigure}[t]{0.3\textwidth}
    \includegraphics[width=\linewidth, valign=t]{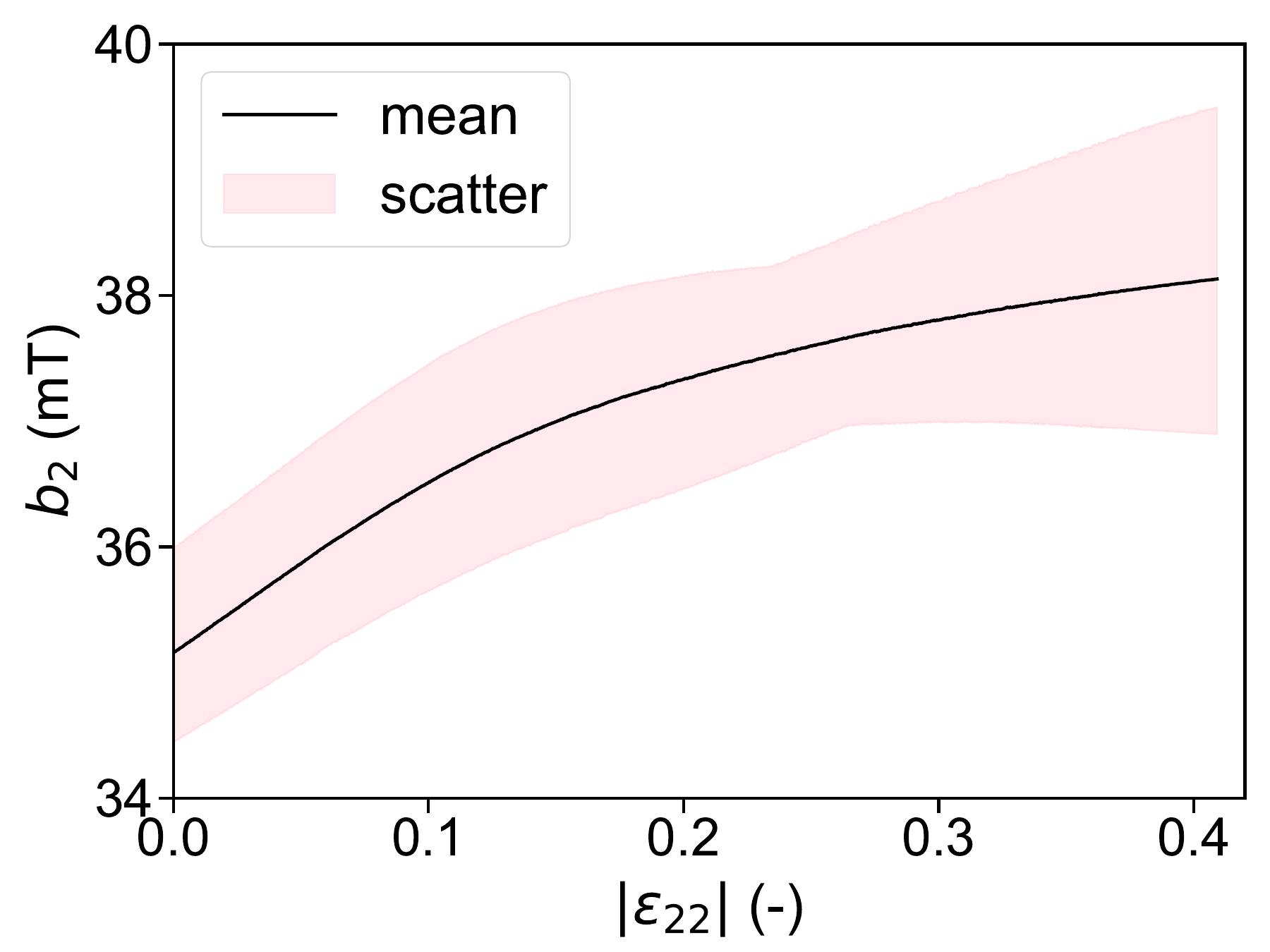}
  \end{subfigure}\hfill
  \begin{subfigure}[t]{0.02\textwidth}
    {\small c)}
  \end{subfigure}
  \begin{subfigure}[t]{0.3\textwidth}
    \includegraphics[width=\linewidth, valign=t]{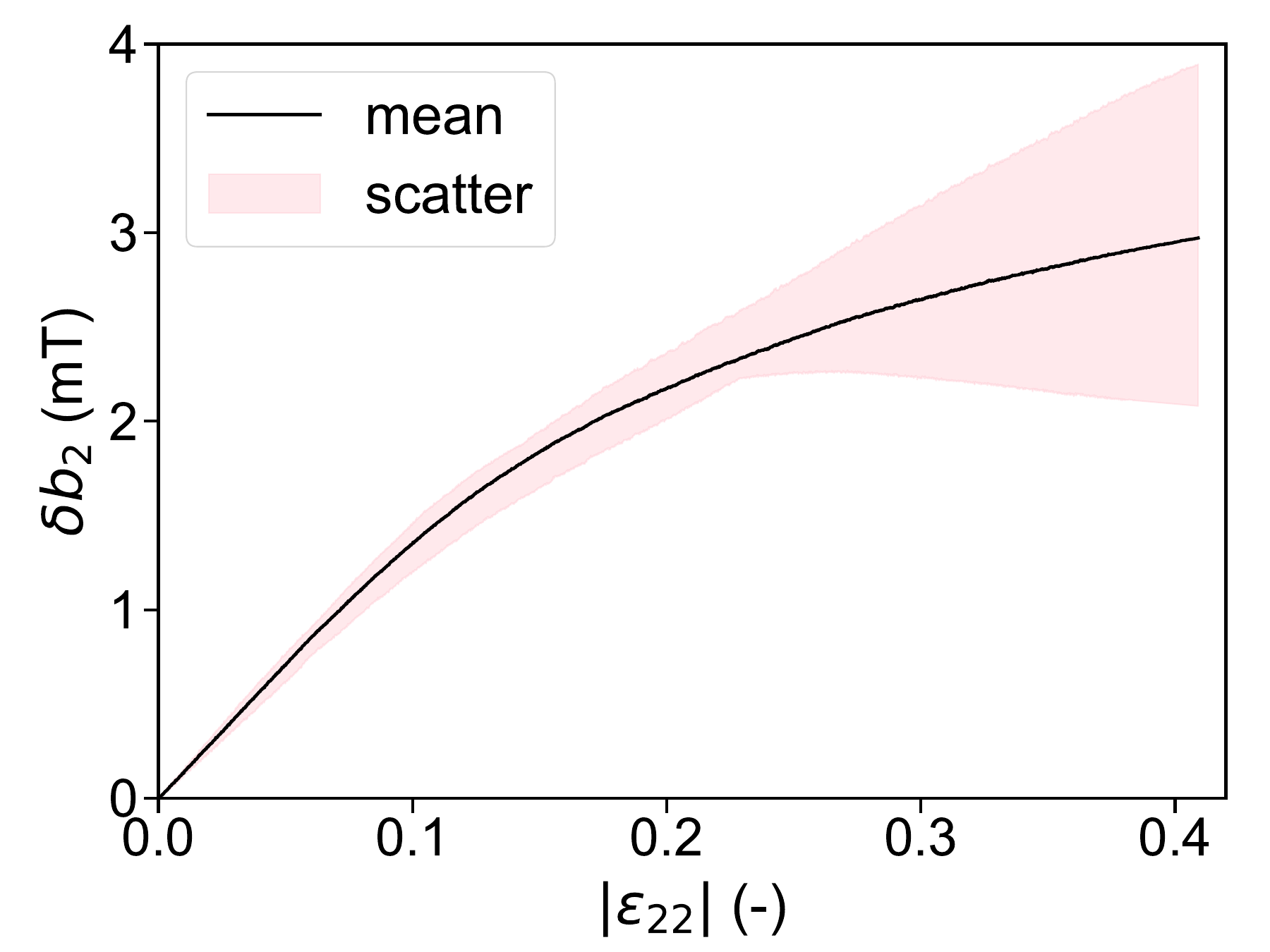}
  \end{subfigure}\hfill
  \caption{Oedometric compression tests for $h$-MRE foams with $c_\pt=0.12$ ($c_\vt=0.45$, $c_\pmt=0.21$). (a) Purely mechanical response of magnetized and unmagnetized cube samples (also tested in all three principal directions). (b) Magnetic flux along the magnetization direction as a function of the applied engineering compressive strain. (c) Variation in the magnetic flux along the magnetization direction as a function of the applied engineering compressive strain.}
  \label{fig:oedometric_experiments}
\end{figure}

Figure~\ref{fig:oedometric_experiments} presents the experimental data from the oedometric compression test. In Fig.~\ref{fig:oedometric_experiments}a, we observe that the mean responses of both the magnetized and unmagnetized samples almost overlap, indicating that the magnetic interaction forces between the particles play a minor role in the mechanical response of the present foam. Moreover, the scatter due to different samples and compression directions is fairly weak even at very large strains, thereby indicating that the mechanical response of the present $h$-MRE foams is fairly isotropic. 

Figures~\ref{fig:oedometric_experiments}b and c show the evolution of the magnetic flux and its variation during the mechanical compression as measured by the Hall sensor at the bottom of the sample. We observe that the compression of the $h$-MRE leads to significant magnetic flux changes (in the order of $\sim3~\si{\milli\tesla}$) given an initial remanent magnetic flux of about $35~\si{\milli \tesla}$, i.e, about 10\% change. These changes are well measured by our Hall sensor that has an accuracy of $\pm 100~\si{\micro\tesla}$. Furthermore, we observe that the initial $\db_2$ change is almost linear followed by a nonlinear response but monotonically increasing. The scatter of $b_2$ appears more significant than that of the stresses but $\db_2$ exhibits scattering only at larger strains. One should keep in mind, however, that the magnetic field sensitivity is expected to be higher than the mechanical one. More discussion on the nature of this data is carried out with the aid of the modeling in Section~\ref{sec:Model calibration and predictive capabilities by comparison with the experiments}. 

\begin{figure}[h!]
  \centering
  \begin{subfigure}[t]{0.02\textwidth}
    {\small a)}
  \end{subfigure}
  \begin{subfigure}[t]{0.3\textwidth}
    \includegraphics[width=\linewidth, valign=t]{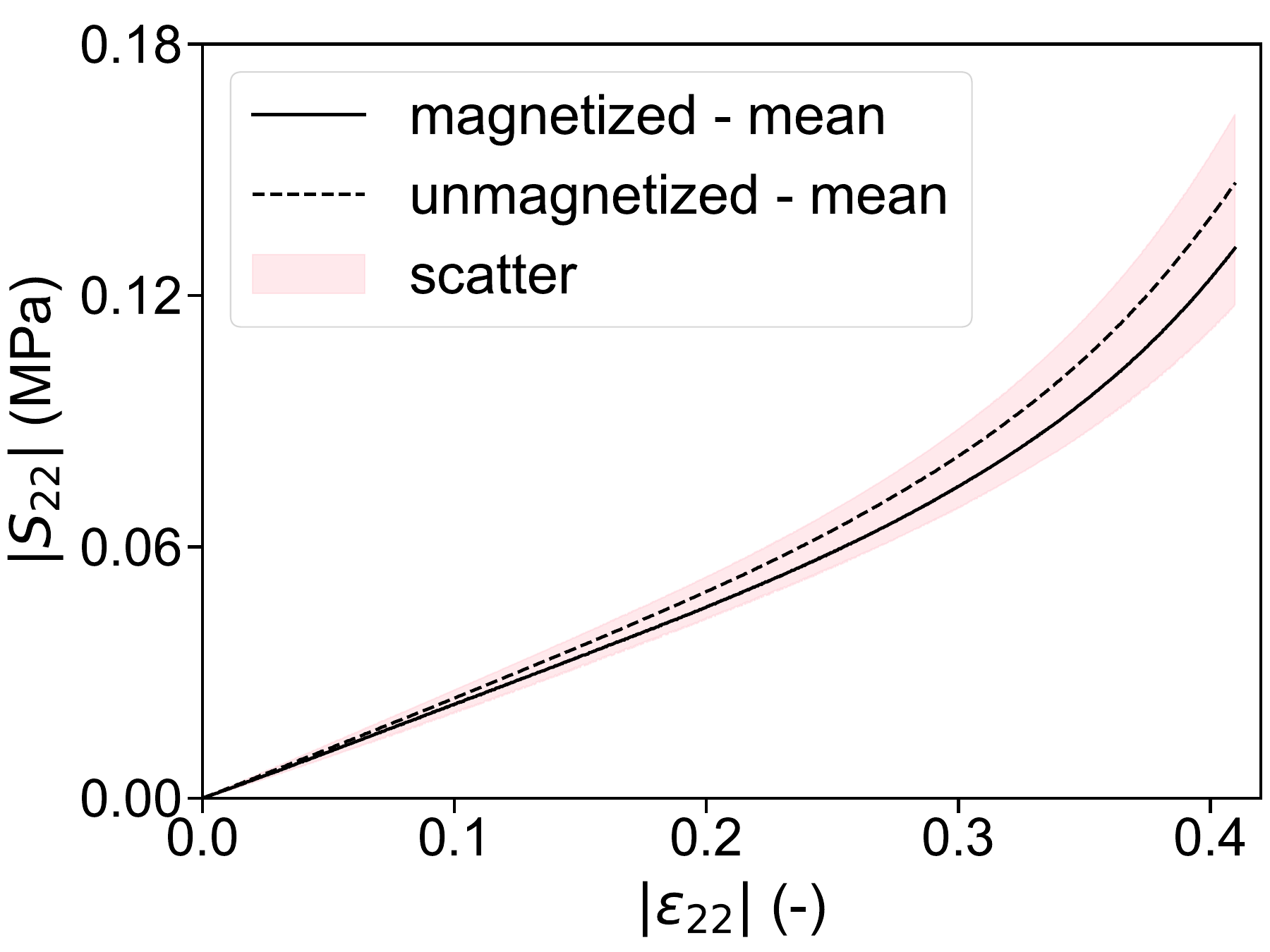}
  \end{subfigure}\hfill
  \begin{subfigure}[t]{0.02\textwidth}
    {\small b)}
  \end{subfigure}
  \begin{subfigure}[t]{0.3\textwidth}
    \includegraphics[width=\linewidth, valign=t]{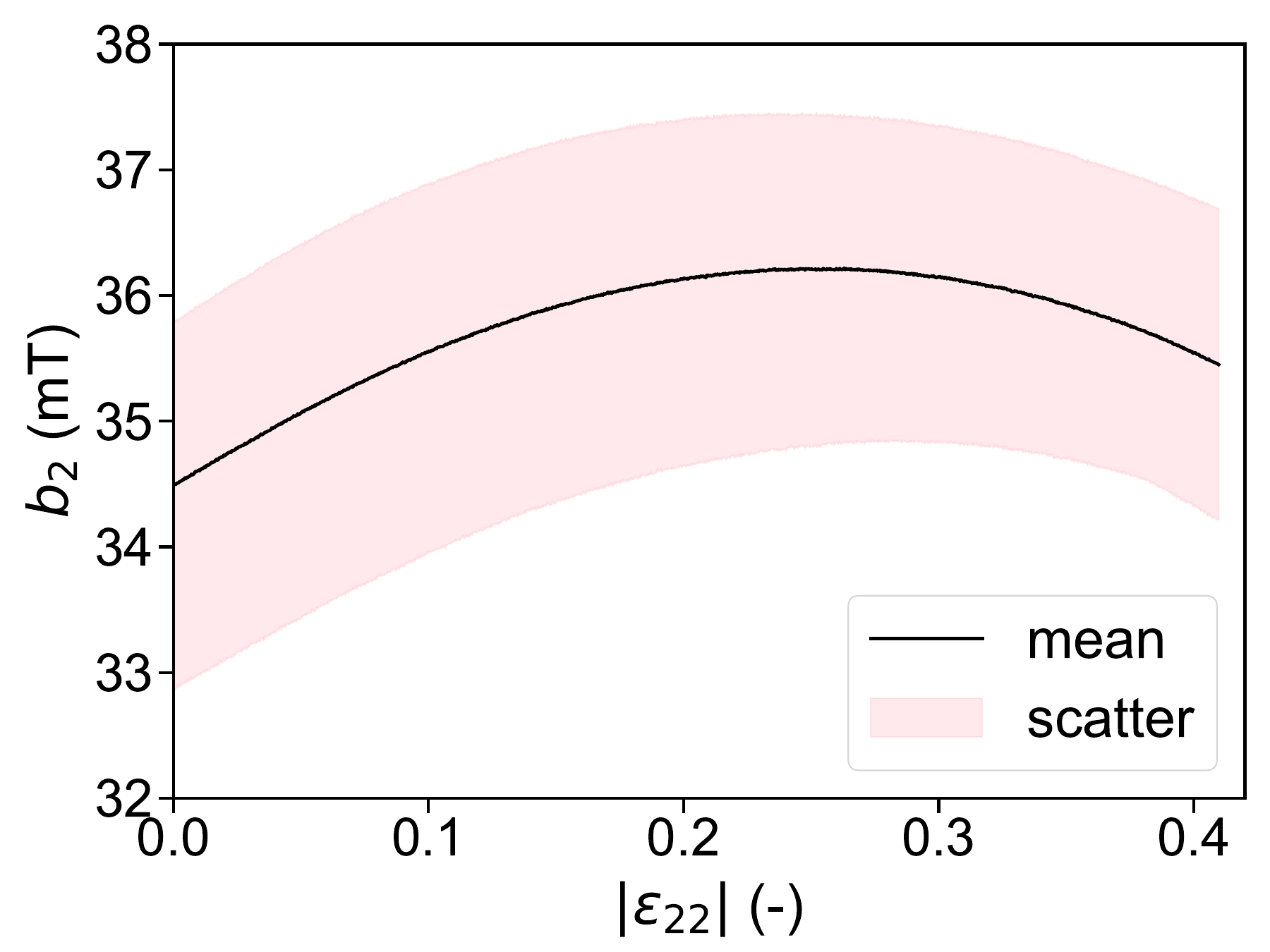}
  \end{subfigure}\hfill
  \begin{subfigure}[t]{0.02\textwidth}
    {\small c)}
  \end{subfigure}
  \begin{subfigure}[t]{0.3\textwidth}
    \includegraphics[width=\linewidth, valign=t]{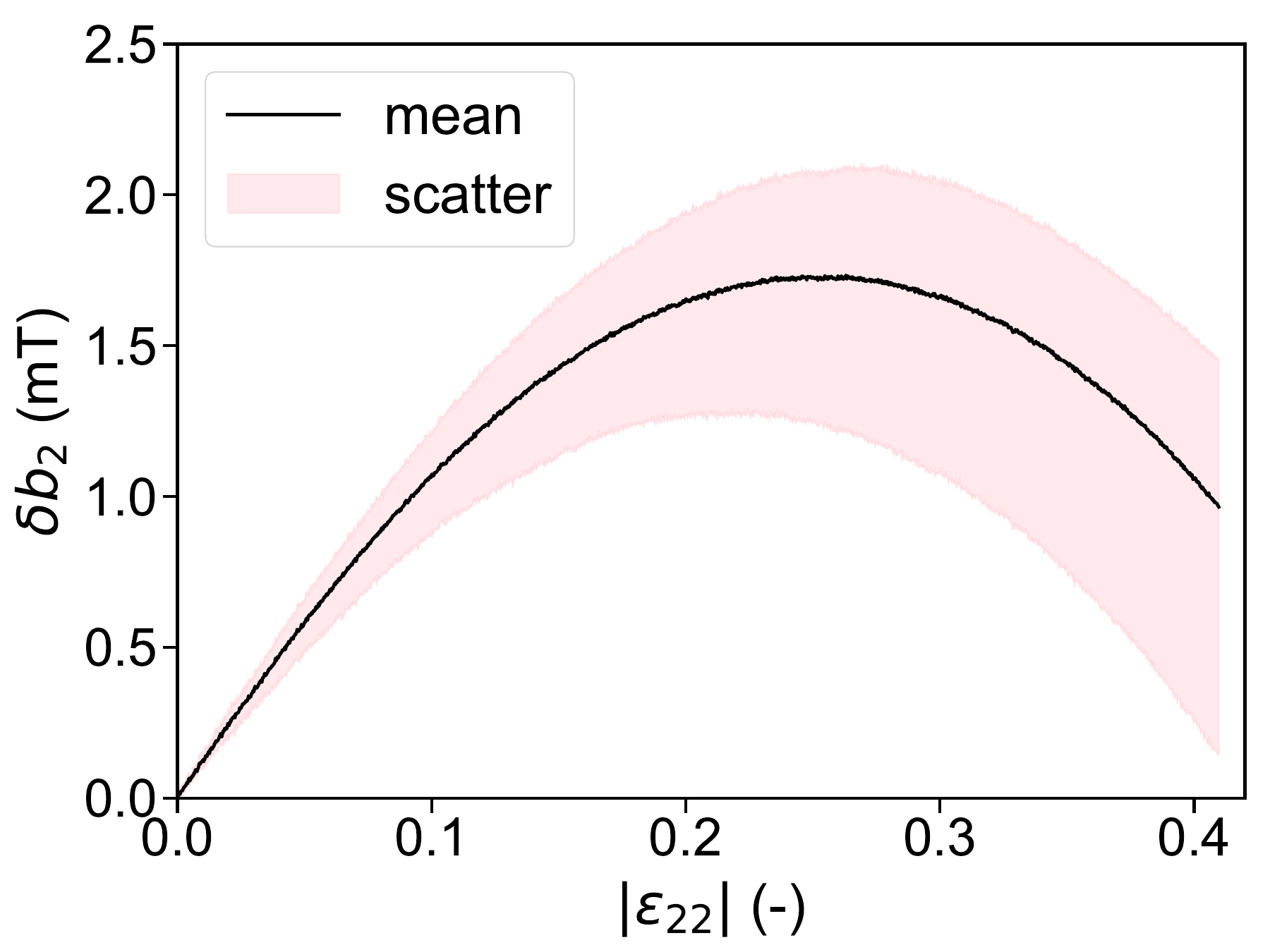}
  \end{subfigure}\hfill
  \caption{Uniaxial compression tests for $h$-MRE foams with $c_\pt=0.12$ ($c_\vt=0.45$, $c_\pmt=0.21$). (a) Purely mechanical response of magnetized and unmagnetized cube samples (also tested in all three principal directions). (b) Magnetic flux along the magnetization direction as a function of the applied engineering compressive strain. (c) Variation in the magnetic flux along the magnetization direction as a function of the applied engineering compressive strain.}
  \label{fig:clamped_experiments}
\end{figure}
Fig.~\ref{fig:clamped_experiments} shows the experimental data from the uniaxial compression stress with traction-free lateral sides. In Fig.~\ref{fig:clamped_experiments}a, we observe again that the mean response of the magnetized and unmagnetized samples are fairly close albeit the magnetized one exhibits a lower stress value at larger strains. Intuitively, this is expected since as the particles come closer together they attract, which aids the material to compress with a lower overall compressive stress. Nevertheless, that difference is still fairly small and will not be taken into account in the subsequent modeling approach for simplicity. Note that shall the foam be even softer, one should consider eventually those differences in the modeling approach \citep{Danas2025}. In turn, the magnetic flux $b_2$ and its variation $\db_2$ in Figs.~\ref{fig:clamped_experiments}b and c, respectively, exhibit a strongly non-monotonic, nonlinear response revealing clearly the complexity of the magneto-mechanical response of the present foams in this configuration. 

\begin{remark}
As we will clarify below with the aid of the magneto-mechanical model and the full-field numerical solution of the relevant boundary value problem (BVP), the observed nonlinear responses in both experiments are a consequence of a non-trivial coupling between the sample shape change and the material volume change due to void closure. In a nutshell, void closure leads to volume decrease and thus increase of the particle concentration in the foam, and \textit{vice versa}. This, in turn, leads to an increase in the effective magnetization of the sample and consequently in the magnetic flux at the bottom face. By contrast, the same compression load induces a decrease in the axial dimension of the sample itself (and elongation along the lateral directions in the traction-free test) leading to a decrease in the magnetic flux at the bottom face (see purely magnetic results in Appendix of \cite{Lin2025}). Those two competing effects come into play in different ways in those two experiments and produce very different magnetic responses. This variability in the measured magnetic response is in fact a highly desirable feature in the present foams---which is much less present in the more conventional \textit{incompressible} $h$-MREs---since it allows to distinguish among various multi-axial deformation states as we detail more in Section~\ref{sec:Numerical study: multi-loading sensing capabilities of hMRE foam}. 
\end{remark}

\section{Thermodynamics, constitutive modeling and calibration}
\label{sec:Thermodynamics, constitutive modeling and calibration}

\subsection{Internal magnetic variable}
\label{subsec:Internal magnetic variable}

In order to model the dissipative character of the magnetic response of the particles, we follow \cite{mukherjee2021} and \cite{mukherjee2022}, and introduce an internal remanent magnetic vector variable 
\begin{align}
	\Hvbr\in\mathbb{R}^3,
\end{align}
which lies in the \textit{stretch-free}, \textit{intermediate} configuration $\mathcal{V}_i$. This internal variable is a remanent H-like vectorial quantity, and the central assumption behind this ``choice'' is that it is not affected by macroscopic stretches but only by rotations. This has been shown to be an accurate assumption for incompressible $h$-MREs in \cite{mukherjee2021} and will also be shown to be a reasonable assumption in the present case of foams too, despite the compressibility of the foam. A plausible explanation lies in the fact that magnetic dissipation takes place at the particle scale and is not (or only very weakly) affected by particle interactions or mechanical strains. This assumption requires further validation via micromechanics calculations and additional experiments but this is beyond the scope of the present work, which focuses mainly on the magneto-mechanical response of an $h$-MRE foam after pre-magnetization.

Note, however, that the above assumption does not imply that the resulting magnetization is independent of strains and volume changes, as was the case in incompressible $h$-MREs (see extensive discussion in \cite{DanasReis2024}). On the contrary, we will show below that in the present case of an $h$-MRE foam the magnetization response is a strong function of the volume changes, i.e., $J=\det \Fb$.

\subsection{Power balance and governing equations}
\label{subsec:Power balance and governing equations}

Considering isothermal conditions, we define an energy density function that depends on the deformation gradient $\Fb$, the magnetic flux $\Bvb$, and the internal variable via $\Hvbr$, denoted as $W\left(\Fb , \Bvb , \Hvbr  \right)$. The local dissipation density $\Diss$ is given as the difference of the external power $\mathcal{P}$ minus the rate of change of the internal energy $\dot{W}$, i.e., \citep{coleman1974,rambausek2021,lucarini2022}
\begin{align}
    \label{eq:dissipation_inequality}
    \Diss = \mathcal{P} - \dot{W} =
        \left(\Svb - \dfrac{\p W}{\p \Fb}\right) \cdot \dot{\Fb} + 
        \left(\Hvb - \dfrac{\p W}{\p \Bvb}\right) \cdot \dot{\Bvb}
        - \dfrac{\p W}{\p \Hvbr} \cdot \Hvbrdot\geq 0.
\end{align}
Owing to the arbitrariness of $\dot{\Fb}$ and $\dot{\Hvb}$, we employ the standard Coleman--Noll--Gurtin argument to arrive at the well-established constitutive relations
\begin{align}
    \label{eq:S-H_constitutive_relations}
    \Svb = \dfrac{\p W}{\p \Fb}, \qquad \Hvb = \dfrac{\p W}{\p \Bvb}.
\end{align}
The last term in equation \eqref{eq:dissipation_inequality} establishes the consistency conditions allowing to define the evolution of $\Hvbr$ by use of the generalized standard material (GSM) \citep{halphen1975} formalism, such that 
\begin{equation}
	\Bvbr= \dfrac{\partial D}{\partial \Hvbrdot}= - \dfrac{\partial W}{\partial \Hvbr}.
	\label{eq:GSM_relation}
\end{equation}
In this last expression, we have introduced for convenience the conjugate variable to $\Hvbr$, denoted for consistency with $\Bvbr$, which is fairly similar to the backstress in purely mechanical elasto-plastic systems.

The thermodynamic conditions presented in this section, together with the field equations and interface/boundary conditions introduced in Section~\ref{sec:Kinematics and magnetostatics}, define the entire set of equations that need to be considered to solve a general boundary value problem in dissipative magneto-elasticity. An incremental variational formulation of the problem may also be constructed based on these local sets of equations as discussed in similar problems in \cite{rambausek2022} and \cite{Danas2024} but is not repeated here for the sake of conciseness.

\subsection{Energy density and dissipation potential}

The aforementioned definition of the independent variables $\Fb$ and $\Bvb$ as well as the internal variable $\Hvbr$ provide infinite possibilities to construct a constitutive model for $h$-MRE foams. Motivated by previous studies in incompressible $h$-MREs and the present experimental results, we propose below only a small subset of such options starting with the definition of the corresponding mechanical invariants
\begin{align}
\label{eq:mech_invars}
    I_1 = \text{tr}(\Cb), \qquad I_3 = J^2 = \det \Cb,
\end{align}
and magneto-mechanical invariants
\begin{align}
\label{eq:FB_invariants}
I_5^{\mathtt{B}} = \Bvb \cdot \Cb \Bvb , \qquad
I_5^{\mathtt{BHr}} = \Bvb \cdot \Cb^{1/2}\Hvbr, \qquad 
I_5^{\mathtt{Hr}} = \Hvbr \cdot \Hvbr.
\end{align}
The energetic invariant $I_5^{\mathtt{B}}$ is usually employed in the context of magnetically-soft MREs \citep{ponte-castaeda2011,danas2017effective,mukherjee2020}. The remaining invariants are mixed or purely remanent ones, and are necessary in the modeling of $h$-MREs in general. Moreover, it is emphasized that $\Hvbr$ has the same units as the h-field and thus proper addition of $\muo$ is required in the final energy expressions. Finally, in order to satisfy the invariance of the dissipation potential, we additionally employ the Euclidean norm $|\Hvbrdot| = \sqrt{\Hvbrdot \cdot \Hvbrdot}$ as an invariant of $\Hvbrdot$.

We consider next the energy function to be the sum of a mechanical and magnetic free energy density, such that
\begin{align}
W(\Cb,\Bvb,\Hvbr) &= \Wmech(I_1,J) + \Wmag(J,I_5^{\mathtt{B}},I_5^{\mathtt{BHr}}, I_5^{\mathtt{Hr}}) + \dfrac{1}{2 \mu_0 J}  I_5^{\mathtt{B}}. \label{eq:WB_full}
\end{align}
In both expressions, $\rho_\foam$ is the \emph{reference} density of the foam, while the last term $I_5^{\mathtt{B}}/2\mu_0$ in \eqref{eq:WB_full} represents the energy associated with free space with $\mu_0$ being the magnetic permeability in vacuum or in non-magnetic solids such as in a polymer matrix phase.

\begin{remark}
Contrary to earlier studies in $h$-MREs, the present experimental data indicate negligible particle-particle interactions and thus intrinsic Maxwell stress effects (see discussion on the stress response for the magnetized and unmagnetized samples in Figs.~\ref{fig:oedometric_experiments} and \ref{fig:clamped_experiments}). This is directly related to the stiffness of the surrounding silicone phase in the present study, as well as the fact that the relative permeability of the pre-magnetized foam is close to unity and thus almost identical to that of the surrounding air. These observations, all together, allow to simplify the identification of the magneto-mechanical constitutive law by ignoring contributions from additional magneto-mechanical invariants in \eqref{eq:WB_full} (see \cite{mukherjee2021}). Yet, we expect that additional invariants will be necessary shall a softer foam be manufactured (see for instance recent works on very soft MREs by \cite{moreno2021,lucarini2022,Garcia2022,Gebhart2022b,Gonzalez2024,Danas2025}). 
\end{remark}

\subsubsection{Mechanical energy density}
\label{subsubsec:Mechanical energy density}

The mechanical energy contribution is highly non-trivial at finite strains for a foam material. In the present study, we propose a two-scale homogenization model based on a combination of two distinct earlier results. We start by defining the mechanical energy density of the foam, considering that it is made up of two phases: an incompressible nonlinear elastic material (which corresponds to the composite $h$-MRE phase itself), denoted by the energy density $\Wmechmat$, and voids (see graphical sketch in Fig.~\ref{fig:molds}). Following the recent works of \cite{Shrimali2019} and \cite{Luo2023}, we write the energy density of the foam as
\begin{equation}
	\Wmech(I_1,J)=(1-c_\vt)\Wmechmat\left(\frac{G_\foam\,(I_1 -3)+\mathcal{F}(J)}{1-c_\vt}+3\right).
	\label{eq:Wmech_foam_def}
\end{equation}
In this expression, $c_\vt$ is the initial reference porosity defined in \eqref{eq:cv_cpm_cp_defs}, $G_\pmt$ is the shear modulus of the incompressible $h$-MRE phase (comprising the polymer matrix and the particles) to be defined below and $\mathcal{F}(J)$ is given by 
\begin{equation}
\mathcal{F}(J)=\frac{3}{J^{1/3}}\left[\mathcal{C}\,(2 J-1)-\mathcal{D}\frac{(1-c_\vt)J^{1/3}(3J^{2/3}+2c_\vt)}{3+2c_\vt}-\mathcal{E}\frac{c_\vt^{1/3}J^{1/3}(2J+c_\vt-2)}{(J-1+c_\vt)^{1/3}}\right],
\end{equation}
with 
\begin{align}
    &\mathcal{C} = \frac{(24+c_\vt-10c_\vt^2)G_\foam-3c_\vt(15+4c_\vt)\kappa_\foam}{(c_\vt-1)(36+31c_\vt)}, \nonumber\\[0.5cm]
    &\mathcal{D} = \frac{(3+2c_\vt)((8+3c_\vt)G_\foam-15c_\vt \kappa_\foam)}{(c_\vt-1)(36+31c_\vt)},\\[0.5cm]
    &\mathcal{E} = \frac{3c_\vt((3+2c_\vt)G_\foam-(9+10c_\vt)\kappa_\foam)}{(c_\vt-1)(36+31c_\vt)}\nonumber.
\end{align}
Hereby, $G_\foam$ and $\kappa_\foam$ denote the normalized effective linear elastic moduli of the foam and can be given by any available homogenization or phenomenological estimate. In this work, we find that the well-known Hashin-Shtrikman estimates \citep{Hashin1963,Willis1983} obtained for porous materials with an incompressible matrix phase are sufficiently accurate, thus taking the simple form
\begin{equation}
G_\foam= \frac{3(1-c_\vt)}{3+2c_\vt}, \qquad \kappa_\foam = \frac{4(1-c_\vt)}{3c_\vt}.
\label{eq:Gfoam_kfoam_def}
\end{equation}
Depending on the type of porosity in the foam (e.g., open, closed, ellipsoidal, connected or disconnected, Gaussian, etc), one may elaborate further on those estimates as discussed in \cite{Zerhouni2021} and \cite{Luo2023}. 

It remains to define the energy density $\Wmechmat$ and the effective shear modulus $G_\pmt$ of the incompressible $h$-MRE matrix phase in the above expressions. For this, we resort to the earlier work of \cite{LopezPamies2013} and write a single-term energy function of the form \citep{LopezPamies2010}
\begin{equation}
\Wmechmat(y)=\frac{3^{1-\alpha}}{2\alpha}G_\pmt\,\left(y^{\alpha}-3^{\alpha}\right), \quad \alpha \in \mathbb{R},
\label{eq:Wmech_hMRE_def}
\end{equation}
where we have used the argument $y$ to avoid confusion with the  use of $I_1$ in \eqref{eq:Wmech_foam_def}, and $G_\pmt$ is given by the approximate linear elastic estimate proposed in \cite{Lefevre2022} (see also \cite{Luo2023})\footnote{In the papers of \cite{Lefevre2022} and \cite{Luo2023}, the full expression includes also a power coefficient (not shown here for brevity). That power coefficient serves to model the percolation limit in the case of mono- or polydisperse particles. In our case, the magnetic particles have a range of sizes and we assume for simplicity that the percolation limit is at $c_\pmt=1$ and thus the corresponding power coefficient takes the value of unity.}
\begin{equation}
	G_\pmt=G_\sil \left[\left(1+c_\pmt^4\right) \left(1-c_\pmt\right)\right]^{-5/2}.
	\label{eq:Gpm_def}
\end{equation}
Here, $G_\sil$ serves to denote the shear modulus of the incompressible cured silicone, whereas $G_\pmt \to \infty$ as $c_\pmt \to 1$ since it corresponds to a homogenization estimate for mechanically rigid particles. Other estimates for $G_\pmt$ may also be used if required by microstructural considerations such as particle poly- or monodispersity and their distributions (see relevant discussion in \cite{Luo2023}).

\begin{remark}
The expressions introduced in this section include a total of two unknown parameters, $G_\sil$ and the power $\alpha$ that will be calibrated in Section~\ref{sec:Model calibration and predictive capabilities by comparison with the experiments} by use of the purely mechanical experiments discussed in Section~\ref{sec:Fabrication and experiments}. Furthermore, the mechanical energy for the foam involves also two microstructural parameters measured directly during fabrication and post-processing of the manufactured foam samples, the porosity $c_\vt$ and the particle volume fraction in the uncured silicone matrix $c_\pmt$, reported in Table~\ref{tab:microstructural_params}. Using relation \eqref{eq:cv_cpm_cp_defs}, these two last experimental estimates provide the total particle volume fraction in the foam of $c_\pt$.
\end{remark}

\subsubsection{Magnetic energy density}

The definition for the magnetic energy $\Wmag$ is inspired by the earlier work of \cite{mukherjee2022} as well as the recent work on the purely magnetic response of $h$-MRE foams by \cite{Lin2025}. Unlike those earlier works where the particle volume fraction remained fixed, here, we need to take into account the total particle concentration evolution due to volumetric changes via the introduction of the current particle volume fraction $\wtilde{c}_\pt=J^{-1} c_\pt$ defined in \eqref{eq:cv_cp_cpm_current_def}$_3$. 

Specifically, we write the magnetic energy as
\begin{align}
    \Wmag(J,I_5^{\mathtt{B}}, I_5^{\mathtt{BHr}}, I_5^{\mathtt{Hr}}) = \Wmag^e(J,I_5^{\mathtt{B}})+\Wmag^r(J, I_5^{\mathtt{BHr}}, I_5^{\mathtt{Hr}}),
     \label{eq:Wmag_def}
\end{align}
where
\begin{align}
    \Wmag^e(J,I_5^{\mathtt{B}}) = -\frac{1}{2\mu_0}\frac{\chi^e}{1+\chi^e}I_5^{\mathtt{B}} ,
     \label{eq:Wmage_def}
\end{align}
and 
\begin{align}
     \Wmag^r(J, I_5^{\mathtt{BHr}}, I_5^{\mathtt{Hr}}) &= \frac{J^{-1}}{1+ s_1(J-1) + s_2 (J-1)^2 } I_5^{\mathtt{BHr}}+ \frac{\mu_0}{2} \left(\chi^e+\frac{1+2 \wtilde{c}_\pt}{3 \wtilde{c}_\pt} \right) I_5^{\mathtt{Hr}} + \frac{\mu_0}{\wtilde{c}_\pt} \frac{(m^s)^2}{\chi^r_\pt} f_\pt \left(\frac{\sqrt{ I_5^{\mathtt{Hr}}}}{J m^s} \right).
     \label{eq:Wmagr_def}
\end{align}
In this last expression, $\chi^r_\pt$ is related to the remanent susceptibility of the particles (see discussion in \cite{mukherjee2021}) and $f_\pt$ denotes any inverse sigmoid function (see \cite{mukherjee2021}), while for simplicity in the present study, we choose a perfectly saturating function given by
\begin{equation}
    f_\pt(x) = (x - 1) \operatorname{tanh^{-1}}\left((1-\eta) x\right) + \ln(1 + x), \qquad \eta=10^{-3}, \quad \forall x\in[0,1].
    \label{eq:fp_sat_fcn_def}
\end{equation}
The small parameter, $\eta$, is introduced to avoid numerical singularity when $x=1$ (corresponding to the fully saturated regime). The term multiplying $I_5^{\mathtt{BHr}}$ in \eqref{eq:Wmagr_def} is a function of $J$ with only two free parameters $s_1$ and $s_2$ that will be calibrated in Section~\ref{sec:Model calibration and predictive capabilities by comparison with the experiments} by use of the oedometric experimental data. 

In turn, the effective magnetic susceptibility and magnetization saturation of the $h$-MRE foam, $\chi^e$ and $m^s$, respectively, may be estimated directly. Specifically, in \cite{Lin2025}, it was shown experimentally and theoretically that these two quantities are only functions of the total particle volume fraction $c_\pt$, since the underlying silicone matrix and the voids act as a single non-magnetic medium with magnetic permeability $\mu_0$ and thus in terms of magnetism the foam behaves as a two-phase composite with the particles being the only magnetic constituent. With this observation at hand and following \cite{mukherjee2022}, one has the homogenization estimates
\begin{equation}
\chi^e = \dfrac{3\wtilde{c}_\pt \chi^e_\pt}{3+(1-\wtilde{c}_\pt) \chi^e_\pt}=\dfrac{3c_\pt \chi^e_\pt}{3 J+(J-c_\pt) \chi^e_\pt}, \qquad  m^s = \wtilde{c}_\pt \ m^s_\pt \bigg( \dfrac{1+\chi^e_\pt }{1 + \chi^e} \bigg)=\dfrac{c_\pt m^s_\pt}{J}\dfrac{1+\chi^e_\pt }{1 + \chi^e}.
\label{eq:chie_ms_estimates}
\end{equation}
In this expression, $\chi^e_\pt$  and $m^s_\pt$ are the energetic magnetic susceptibility (see discussion in \cite{mukherjee2021}) and the magnetization saturation of the particles, respectively. 

\begin{remark}
For the full description of the magnetic energy, one has to prescribe the magnetic properties of the particles, $\chi^e_\pt$, $\chi^r_\pt$ and $m^s_\pt$. The parameter $\chi^r_\pt$ affects only the dissipative response of the $h$-MRE foam but not the subsequent mechanical response after pre-magnetization and for small or negligible applied magnetic fields. In turn, in NdFeB magnetic materials $\chi^e_\pt$ is very small \citep{zhao2019,Yan2023}. In view of this and in order to simplify our material parameter identification in the following, we set $\chi^r_\pt=8$ as in the previous studies of \cite{mukherjee2019} and \cite{mukherjee2021}, and assume an ideal magnetic response with $\chi^e_\pt=0$ (which implies also $\chi^e=0$ in \eqref{eq:chie_ms_estimates}). This leads to the simple form for the magnetization saturation of the foam
\begin{equation}
m^s=J^{-1}c_\pt m^s_\pt.
\label{eq:ms_msp_ideal_magnet}
\end{equation}
with $m^s_\pt$ directly identified from the present experimental data as discussed below in Section~\ref{sec:Model calibration and predictive capabilities by comparison with the experiments}.
\end{remark}

\subsubsection{Dissipation potential for an ideal magnet}
\label{subsubsec:Dissipation potential}

We define the dissipation potential, $D$, as a function of the remanent magnetic field rate $\Hvbrdot$ and focus on ideal magnets only ($\chi^e_\pt=0$). We also assume that magnetic dissipation is independent of deformation and only depends on the state of magnetic field in the particles. In order to induce dissipative phenomena (related to magnetic wall motion at the nanometer scale) in each particle, a strong magnetic field is necessary. In earlier works \citep{mukherjee2019,mukherjee2021}, we have shown that the initiation of magnetic switching is almost independent of the particle volume fraction indicating that it is mainly driven by the externally applied magnetic field, whereby interparticle interactions have only a negligible effect on it. In turn, as presented above, magnetic saturation is a linear function of the particle volume fraction. A consequence of these observations is that one may use exactly the same dissipation potential used for incompressible $h$-MREs. For completeness, we present briefly the main elements of that model. 

Following \cite{mukherjee2021}, we consider that magnetic dissipation is rate-independent (at least for slow load rates as is the case here) and thus one may define the dissipation potential to be a homogeneous convex function of $\Hvbrdot$ of degree one, i.e., 
\begin{equation}
D(\Hvbrdot) = b^c \norm{\Hvbrdot}\ge 0. \label{eq:Dissi_pot_def}
\end{equation}
Here, $\norm{.}$ is the standard Eulerian norm allowing to satisfy material frame indifference and $b^c$ is the effective \emph{coercive field} of the $h$-MRE foam given by
\begin{equation}
b^c = b^c_{\mathtt{p}}, 
\label{eq:bc_def}
\end{equation}
where $b^c_{\mathtt{p}}$ is the particle coercivity and is taken as constant in the context of ideal magnets \citep{idiart2006}. For non-ideal magnetic responses, a correction term may be added as discussed in \cite{mukherjee2021}. An important observation here is that the apparent coercive field for the foam and the $h$-MRE is that of the particle and is independent of the particle volume fraction. This implies directly that changes in the total particle volume fraction due to volumetric changes via $J$ will not affect $\Hvbr$, which is estimated as discussed below. Again, this proposition needs to be confirmed further by numerical homogenization calculations and experiments of the full dissipative response but this is beyond the scope of the present study. A first indication of such a response is provided in the recent work of \cite{Lin2025} and corresponding microstructural calculations carried out therein.  

The dissipation potential in \eqref{eq:Dissi_pot_def} allows to define a \emph{ferromagnetic switching surface}---similar to the yield surface in rate-independent elastoplasticity---as
\begin{equation}
\Phi(\Bvbr):= \norm{\Bvbr}^2 - (b^c)^2 = 0, \label{eq:def_swsurf}
\end{equation}
which must be satisfied during a magnetic loading/unloading cycle. Subsequent use of the normality rule and positive dissipation implies directly the Karush--Kuhn--Tucker (KKT) conditions \citep{Karush1939,Kuhn1951}, i.e.,  
\begin{equation}
\Hvbrdot = \dot{\Lambda} \dfrac{\partial \Phi}{\partial \Bvbr}, \qquad \Phi(\Bvbr) \leq 0, \qquad \dot{\Lambda} \geq 0 \qquad \text{and} \qquad \dot{\Lambda} \Phi = 0,  \label{eq:KKT_FH}
\end{equation}
with $\Hvbr(t=0)=\mathbf{0}$.

\begin{remark}
The proposed dissipation potential serves to estimate the remanent magnetic field $\Hvbr$ during the pre-magnetization phase. This consists in carrying out a half cycle of strong magnetization of the foam beyond saturation and full unloading to zero external magnetic field. Such a process leads to a permanent magnetic foam with a given amplitude and direction of $\Hvbr$. The subsequent application of purely mechanical loads, which is of interest in the present study, does not alter any further $\Hvbr$ (but does alter $m^s$ and thus $\mvb$) and thus does not require to resolve the above KKT conditions and the switching surface constraint in that case. In view of this, in the following section we propose a simplified form of the above equations for a constant vector $\Hvbr$.
\end{remark}

\subsection{Analytical magnetization estimate for ideal magnets after pre-magnetization}

Let us assume a strongly pre-magnetized state of an ideal $h$-MRE foam such that $\Hvbr$ has reached a constant amplitude and orientation. At the end of the pre-magnetization process and with no remanent strains or volume changes as is the case in the previously discussed model, $\Hvbr$ reaches the amplitude $\norm{\Hvbr}=m^s=c_\pt m^s_\pt$ (i.e., $\Hvbrdot=\mathbf{0}$). This value remains constant even when the applied mechanical loads are significant. Therefore, for the estimation of the \textit{magnetic} response all energy terms that involve only $\Hvbr$ (but not $\Bvb$) become inconsequential and can be dropped, including the dissipation potential. 

In this regard, assuming that the pre-magnetization takes place along direction $2$, i.e., $\Hvbr=c_\pt m^s_\pt \ev_2$, we can compute the magnetization response as \citep{danas2017effective}
\begin{equation}
	\mvb=-\Fb^{-\text{T}}\dfrac{\p (\Wmag)}{\p \Bvb}=\dfrac{c_\pt m^s_\pt\,J^{-1}}{1+ s_1(J-1) + s_2 (J-1)^2 }\Fb^{-\text{T}}\Cb^{1/2}\,\ev_2.
	\label{eq:mv_simplified_ideal}
\end{equation}
This result will allow to calibrate analytically the magnetization saturation $m^s_\pt$ and the coefficients $s_1$ and $s_2$ from the oedometric test data, as discussed in the next section.

\section{Model calibration and predictive capabilities by comparison with the experiments}
\label{sec:Model calibration and predictive capabilities by comparison with the experiments}

In the context of the compressive oedometric test discussed in Fig.~\ref{fig:exp_setup}a, the deformation gradient may be assumed of the form (ignoring frictional effects of the side walls)
\begin{equation}
\Fb=\ev_1\otimes\ev_1+\lambda_2\,\ev_2\otimes\ev_2+\ev_3\otimes\ev_3, \qquad \forall\lambda_2\in(1-c_\vt,1], \quad J=\lambda_2, \quad \varepsilon_{22}=\lambda_2-1\leq 0,
\label{eq:F_oedometric}
\end{equation}
where $\varepsilon_{22}\in(-c_\vt,0]$ is the compressive engineering strain in this case. In addition, use of definitions \eqref{eq:cv_cp_cpm_current_def} in the above expressions leads to $\wtilde{c}_\vt\to 0$ as  $\lambda_2\to 1-c_\vt$, thus attaining the theoretical densification strain for the material. Beyond that point, the $h$-MRE foam becomes incompressible and the resulting stress goes to infinity in the oedometric test. Practically, the experiment has to stop sufficiently early to avoid failure of the load cells used to measure the forces. Next, using the result for the current magnetization in equation \eqref{eq:mv_simplified_ideal}, we have in this case that
\begin{equation}
	\mvb=m_2\ev_2, \qquad m_2=c_\pt m^s_\pt \left[\lambda_2  \left(1+s_1\,(\lambda_2 -1) +s_2\,(\lambda_2 -1)^2\right)\right]^{-1}.
	\label{eq:m2_oedometric_sol}
\end{equation}

In order to connect that solution with the experimental data, one needs to take into account the geometry changes of the specimen with increasing compressive load. As the sample geometry evolves from a cube to a square cuboid in the oedometric test, the surrounding magnetic fields measured during the experiments also change (since a different magnet shape induces different fields). In the case of a rectangular (or square) cuboid, \cite{Gou2004} have proposed an analytical solution inside and outside the magnetic body given a uniform magnetization state inside the magnet, which is exactly the case for the induced cuboid geometries in the present oedometric test (see Fig.~\ref{fig:contours_oedometric_clamped} below as well as \cite{Lin2025}). The \cite{Gou2004} solution, however, has been proposed in the context of pure metallic magnets and as such does not involve finite deformations. Yet, provided the magnetization remains \textit{constant} in the solid, one can use it together with the solution in \eqref{eq:m2_oedometric_sol} to provide an evolving solution of the magnetic fields in the surrounding and inside an evolving magnetic foam that retains a square cuboid symmetry. In view of these observations, we report below the solution for the second component of the current magnetic field $b_2$, while the expressions for the remaining components $b_1$ and $b_3$ are given in the~\ref{Appx:Analytical solution of Gou et al.}.

Consider first that the $h$-MRE foam is a cuboid with dimensions $L_1\times L_2\times L_3$ in the reference configuration. Then, given the form of the deformation gradient in \eqref{eq:F_oedometric}, the deformed dimensions of the foam become 
\begin{equation}
   l_1 = L_1, \quad l_2 = \lambda_2 L_2, \quad l_3 =  L_3,
\end{equation}
whereas the deformed position vector is simply
\begin{equation}
\xv =
\begin{cases}
    \Fb \Xv, & \text{if } X_1 \in [0, L_1], \ X_2 \in [0, L_2], \,\, \text{and } X_3 \in [0, L_3], \\
    \Xv, & \text{otherwise}.
\end{cases}
\end{equation}
Then, using \cite{Gou2004}, the current magnetic field along the pre-magnetized direction $2$ reads
\begin{align}
b_2(x_1,x_2,x_3)= \dfrac{\mu_0 \,m_2}{4\pi} \Big[ &F_2(l_1-x_1,l_2-x_2,l_3-x_3) + F_2(l_1-x_1,l_2-x_2,x_3)\nonumber\\[1ex]
+&F_2(x_1,l_2-x_2,l_3-x_3) + F_2(x_1,l_2-x_2,x_3)\nonumber\\[1ex]
+&F_2(l_1-x_1,x_2,l_3-x_3) + F_2(l_1-x_1,x_2,x_3)\nonumber\\[1ex]
+&F_2(x_1,x_2,l_3-x_3) + F_2(x_1,x_2,x_3)\Big],
\label{eq:b2r_Gou}
\end{align}
with $m_2$ given by equation \eqref{eq:m2_oedometric_sol} and $F_2$ being a non-dimensional function of the spatial current coordinates defined as
\begin{align*}
F_2(f_1,f_2,f_3)=
\begin{cases}
\arctan\left[\dfrac{f_2 \sqrt{f_1^2+f_2^2+f_3^2}}{f_1\,f_3}\right], \qquad f_1, f_3 \neq 0\\[2ex]
0, \qquad \text{otherwise}.
\end{cases}
\end{align*}
The above expressions allow to trace anywhere inside or outside the $h$-MRE foam the evolution of the $b_2$ field analytically. This simplifies substantially the calibration of $m^s_p$, $s_1$ and $s_2$ directly from the experimental data. 

\begin{figure}[h!]
  \centering
  \begin{subfigure}[t]{0.02\textwidth}
    {\small a)}\hfill
  \end{subfigure}
  \begin{subfigure}[t]{0.45\textwidth}
    \includegraphics[width=\linewidth, valign=t]{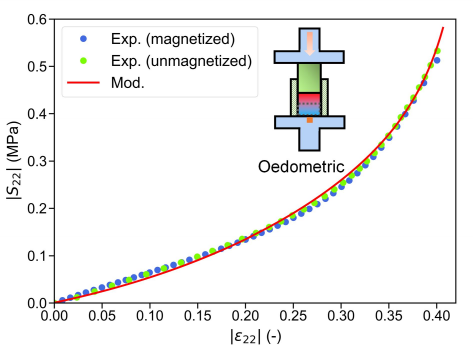}
  \end{subfigure}\hfill
  \begin{subfigure}[t]{0.02\textwidth}
    {\small b)}\hfill
  \end{subfigure}
  \begin{subfigure}[t]{0.45\textwidth}
    \includegraphics[width=\linewidth, valign=t]{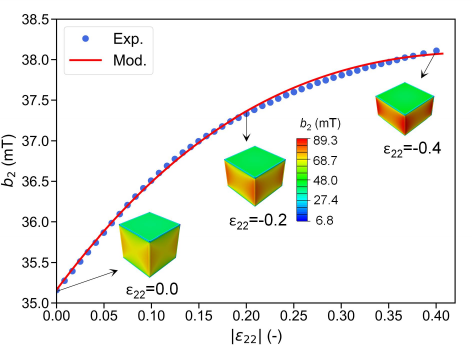}
  \end{subfigure}\\[2ex]
  \begin{subfigure}[t]{0.02\textwidth}
    {\small c)}\hfill
  \end{subfigure}
  \begin{subfigure}[t]{0.45\textwidth}
    \includegraphics[width=\linewidth, valign=t]{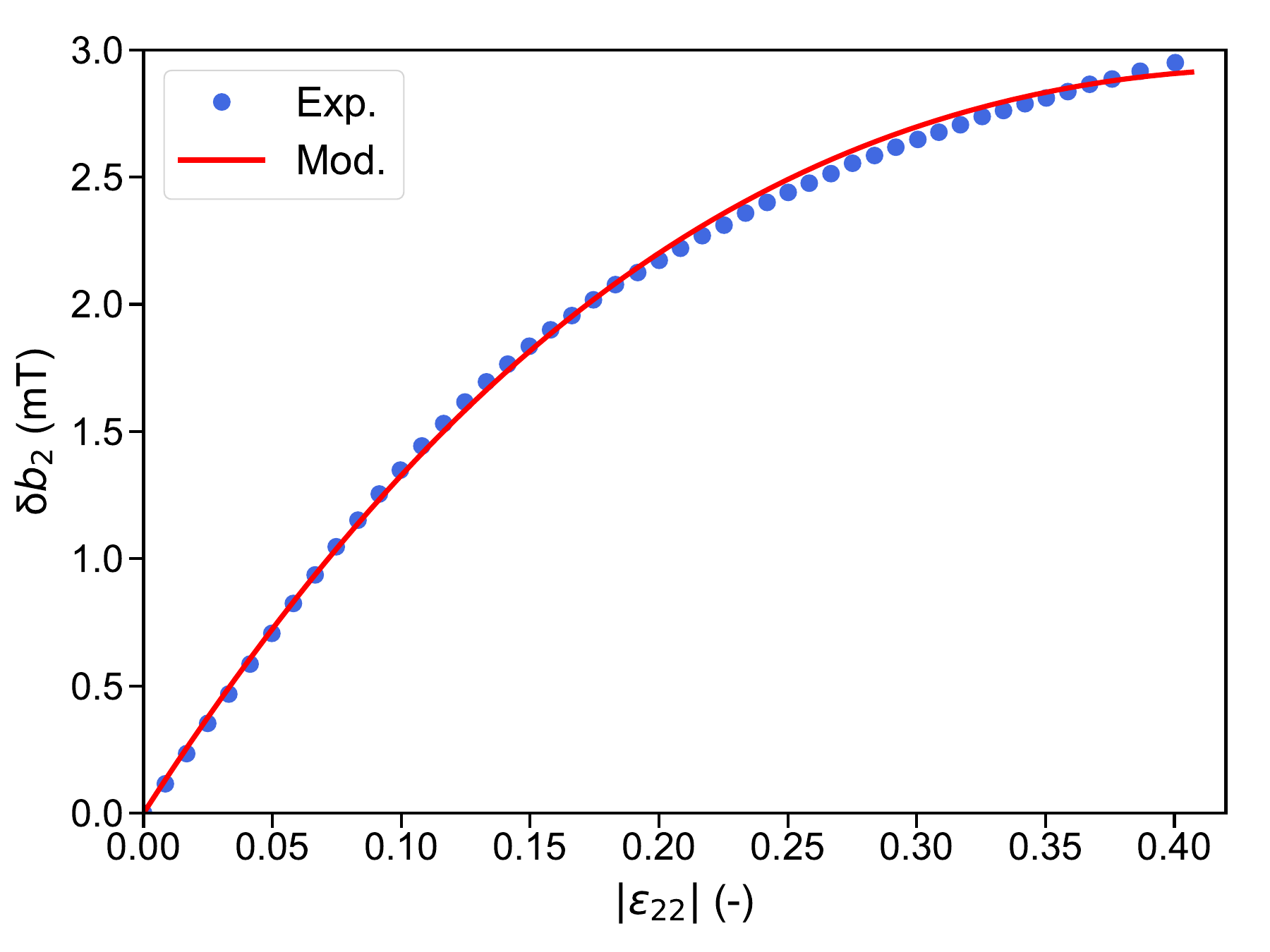}
  \end{subfigure}\hfill
  \begin{subfigure}[t]{0.02\textwidth}
    {\small d)}\hfill
  \end{subfigure}
  \begin{subfigure}[t]{0.45\textwidth}
    \includegraphics[width=\linewidth, valign=t]{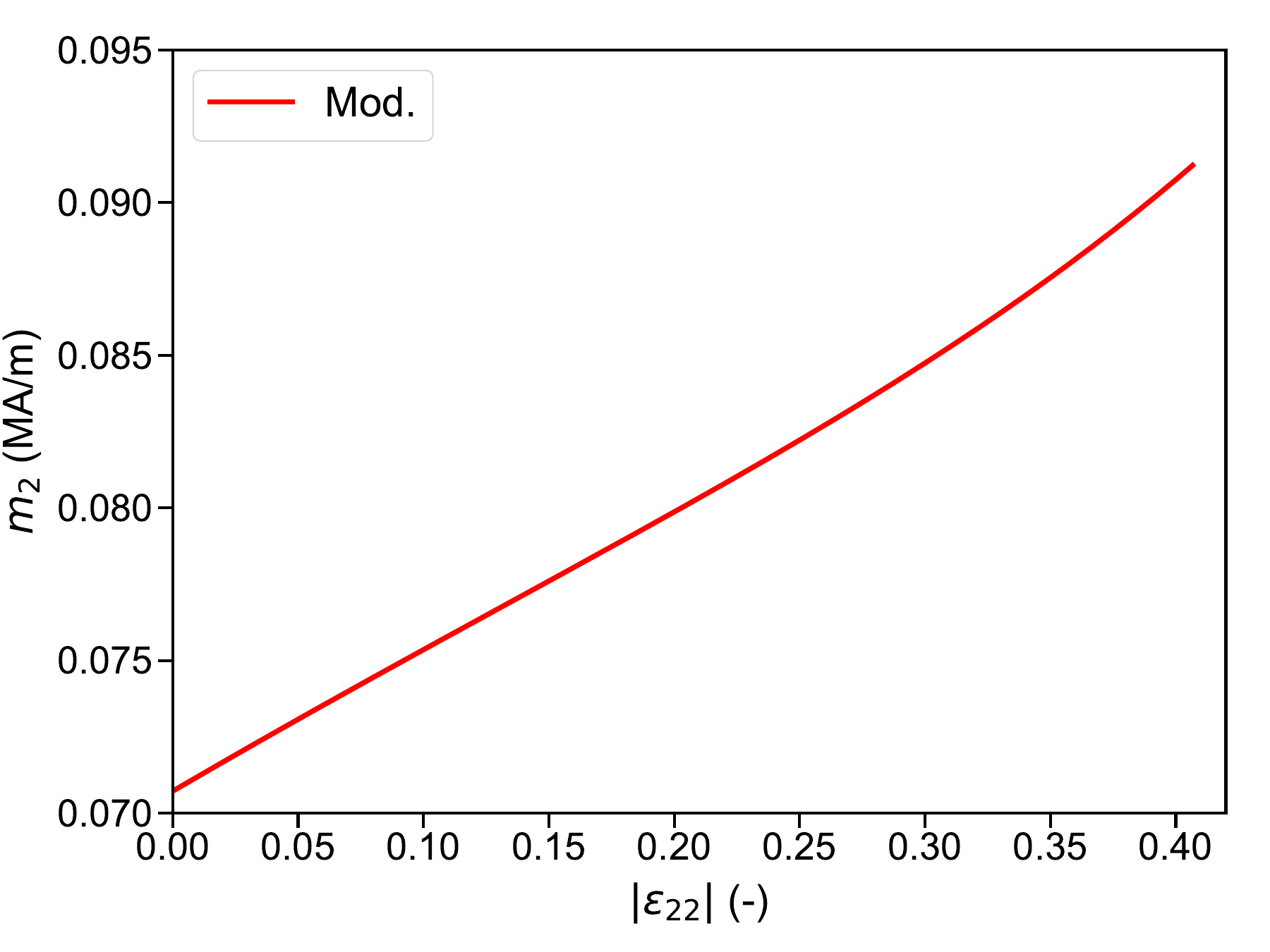}
  \end{subfigure}\hfill
  \caption{Experiments (symbols) versus model (cont. lines) for the magnetized and unmagnetized samples with $c_\pt=0.12$ subjected to oedometric compression conditions. (a) Engineering compressive stress obtained as the measured force divided by the initial cube cross-section (mechanical stress fields are uniform in the model), (b) current magnetic flux and (c) current magnetic flux variation along the magnetization direction measured at the mid-bottom plane of the cube at an offset distance of $-0.5~\si{\milli \meter}$. (d) Magnetization (uniform) in the sample obtained by the model. Numerical simulations and analytical estimates of the model coincide in this case. }
  \label{fig:exp-mod_calibration_oedometric}
\end{figure}
We now have all the ingredients allowing to identify the unknown material constants in our problem. In Table~\ref{tab:MagPropertiesParticles}, we report the magnetic properties used in the calculation of the dissipation response of the $h$-MRE foam. Those properties (with the exception of $\chi^e_\pt$ that was set to zero for simplicity) are directly taken from the earlier studies of \cite{mukherjee2019} and  \cite{mukherjee2021}. Their choices are inconsequential for the subsequent magneto-mechanical response provided that any applied magnetic field is small (in fact, it is zero in the present work), but are given here for completeness.
\begin{table}[h]
\centering
\caption{Known magnetic properties of the NdFeB particles.}
\begin{tabular}{c c c c} \hline
$\chi_{\pt}^e$ & $\chi^r_{\pt}$ & $b^c_{\pt}$ (T) & $\mu_0$ ($\si{\micro\newton}\cdot\mathrm{A}^2$) \\ \hline
0  & 8.0 & 1.062 & $4\pi 10^{-1}$ \\ \hline 
\end{tabular}
\label{tab:MagPropertiesParticles}
\end{table}
\begin{table}[h]
\centering
\caption{Identified mechanical and magnetic properties.}
\begin{tabular}{c c c c c} \hline
$G_\sil$ (kPa) & $\alpha$ & $\mu_0 m^s_{\pt}$ (T) & $s_1$ & $s_2$ \\ \hline
107.16 & -1.22 & 0.765 & -0.321252 & 1.06466 \\ \hline 
\end{tabular}
\label{tab:MechMagProperties}
\end{table}

The identification of the remaining material constants listed in Table \ref{tab:MechMagProperties} is done for the sample with the largest particle concentration, i.e., $c_\pt=0.12$ in the following sequence:
\begin{enumerate}[label={Step \theenumi:},leftmargin=0.7in,resume]

\item The identification of the silicone shear modulus $G_\sil$ and the power exponent $\alpha$ is conducted by first imposing the oedometric deformation gradient defined in \eqref{eq:F_oedometric} and solving for the first Piola\footnote{The evaluation of $S_{22}$ is analytical but too cumbersome to be presented in the text. The relevant Wolfram Mathematica files and data are available upon request.} in the \textit{unmagnetized} case as $S_{22}=\p \Wmech/ \p F_{22}$ with $\Wmech$ defined in \eqref{eq:Wmech_foam_def}. Then, using the \texttt{FindFit} option in Wolfram Mathematica, we calibrate the two constants by use of the experimental data for the unmagnetized specimen, as shown in Fig.~\ref{fig:exp-mod_calibration_oedometric}a.

\item The magnetization saturation of the particles, $m^s_\pt$ is identified\footnote{It is remarked here that the value of $m^s_\pt$ depends on the choice of the saturation function $f_p$ defined in \eqref{eq:fp_sat_fcn_def} and thus is not a universal material property as usually considered in the literature. For instance, in \cite{Lin2025} a different sigmoid function corresponding to a non-ideal magnet was used leading to a slightly lower value for $m^s_\pt$.} using the experimental remanent magnetic field $b_2$ for zero applied strain $\varepsilon_{22}=0$ in Fig.~\ref{fig:exp-mod_calibration_oedometric}b in combination with the \cite{Gou2004} solution \eqref{eq:b2r_Gou} for $J=1$. 

\item The two coefficients $s_1$ and $s_2$, introduced in \eqref{eq:Wmagr_def} are calibrated by employing equation \eqref{eq:b2r_Gou} for $b_2$ as a function of $J$ for the oedometric deformation gradient \eqref{eq:F_oedometric}. The \texttt{FindFit} option in Wolfram Mathematica is also employed in this case.

\end{enumerate}

The calibrated model results are compared with the experimental data in Fig.~\ref{fig:exp-mod_calibration_oedometric}b for $b_2$ and (c) for $\delta b_2(=b_2-b_2^{\varepsilon_{22}=0})$. In turn, Fig.~\ref{fig:exp-mod_calibration_oedometric}d shows the corresponding estimate for $m_2$ defined in \eqref{eq:m2_oedometric_sol} for the corresponding identified constants.

\subsection{Finite element implementation and model assessment}
\label{subsec:Finite element implementation and model assessment}

The model proposed in this work is implemented in a user element (UEL) subroutine (available upon request) in the general purpose finite element (FE) software Abaqus allowing to solve more complex boundary value problems beyond the oedometric test. We further note that the FE implementation of a vector potential formulation requires special attention to make sure the $\Avb$ is uniquely defined at the nodal positions. In the present case, we have implemented the recent $L^2-$norm approach proposed in \cite{Gebhart2024}, which is slightly more convergent than our earlier approach discussed in \cite{dorn2021}, which uses a Coulomb gauge with under-integration. Both approaches deliver indistinguishable results overall in the present case, even though the $L^2-$norm is perhaps easier to implement and leads to slightly less magnetic flux concentrations at the apex of the magnetic cube. For the mesh, we use hexahedral 8-node bilinear elements with full integration and six d.o.f. per node, three for the displacement vector and three for the magnetic vector potential. Furthermore, we use a cubic air box which is ten times larger than that of the $h$-MRE foam sample leading to an approximately $6\times 10^5$~d.o.f. ($10^5$ nodes). One full simulation, including half-cycle pre-magnetization and subsequent mechanical load takes approximately 2--4~h (depending on the boundary conditions) in a 40~cpu cluster. The analytical model curves obtained by using the \cite{Gou2004} solution and the corresponding FE results for the oedometric test coincide exactly in Fig.~\ref{fig:exp-mod_calibration_oedometric} and thus are not distinguished further in those plots. Subsequently, all model results are obtained by use of the FE simulations.

\begin{figure}[h!]
  \centering
  \begin{subfigure}[t]{0.02\textwidth}
    {\small a)}\hfill
  \end{subfigure}
  \begin{subfigure}[t]{0.45\textwidth}
    \includegraphics[width=\linewidth, valign=t]{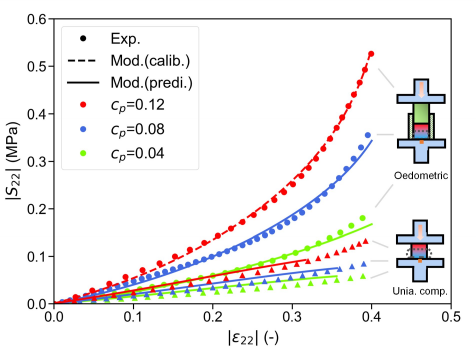}
  \end{subfigure}\hfill
  \begin{subfigure}[t]{0.02\textwidth}
    {\small b)}\hfill
  \end{subfigure}
  \begin{subfigure}[t]{0.45\textwidth}
    \includegraphics[width=\linewidth, valign=t]{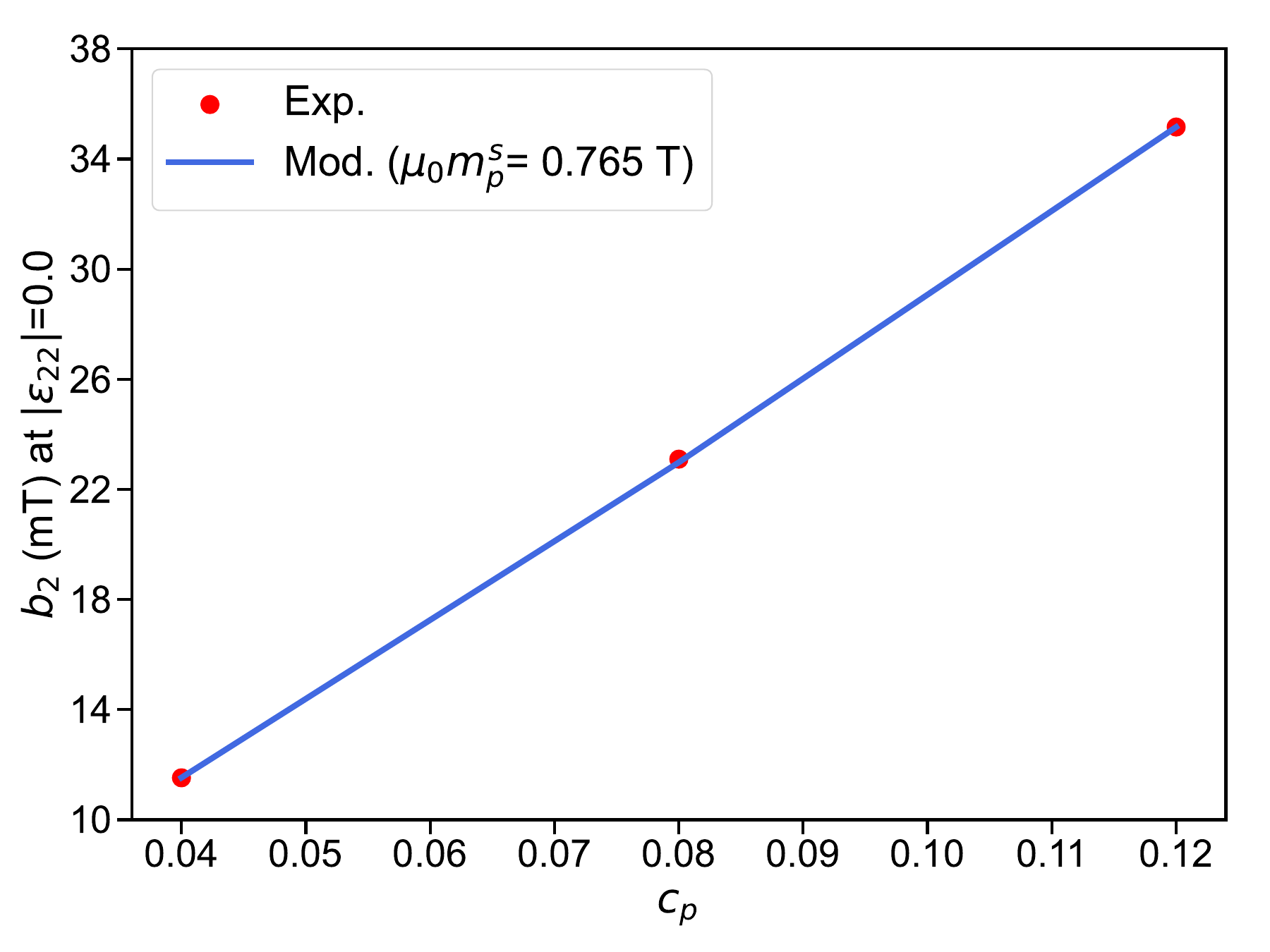}
  \end{subfigure}\\
  \begin{subfigure}[t]{0.02\textwidth}
    {\small c)}\hfill
  \end{subfigure}
  \begin{subfigure}[t]{0.45\textwidth}
    \includegraphics[width=\linewidth, valign=t]{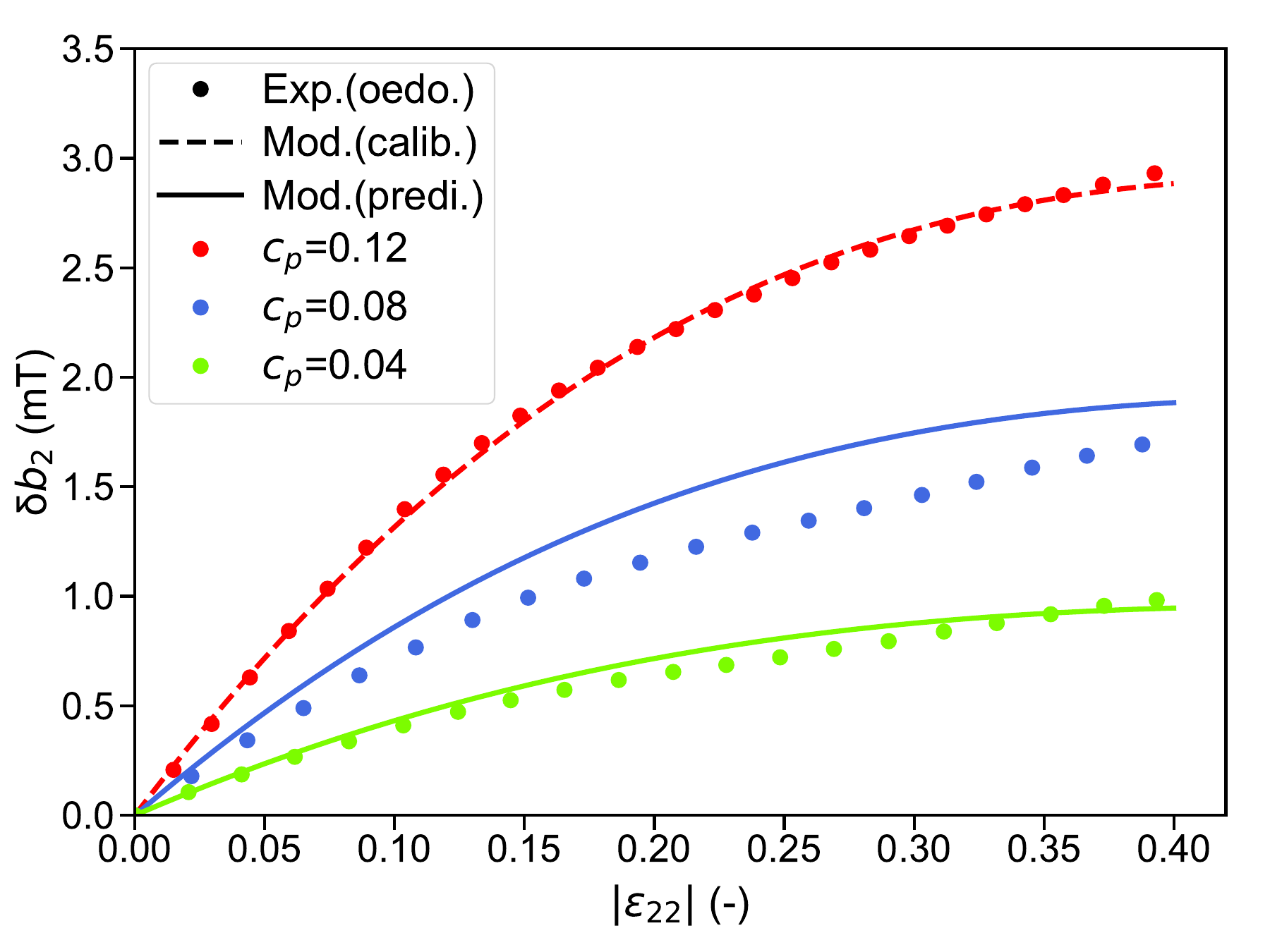}
  \end{subfigure}\hfill
  \begin{subfigure}[t]{0.02\textwidth}
    {\small d)}\hfill
  \end{subfigure}
  \begin{subfigure}[t]{0.45\textwidth}
    \includegraphics[width=\linewidth, valign=t]{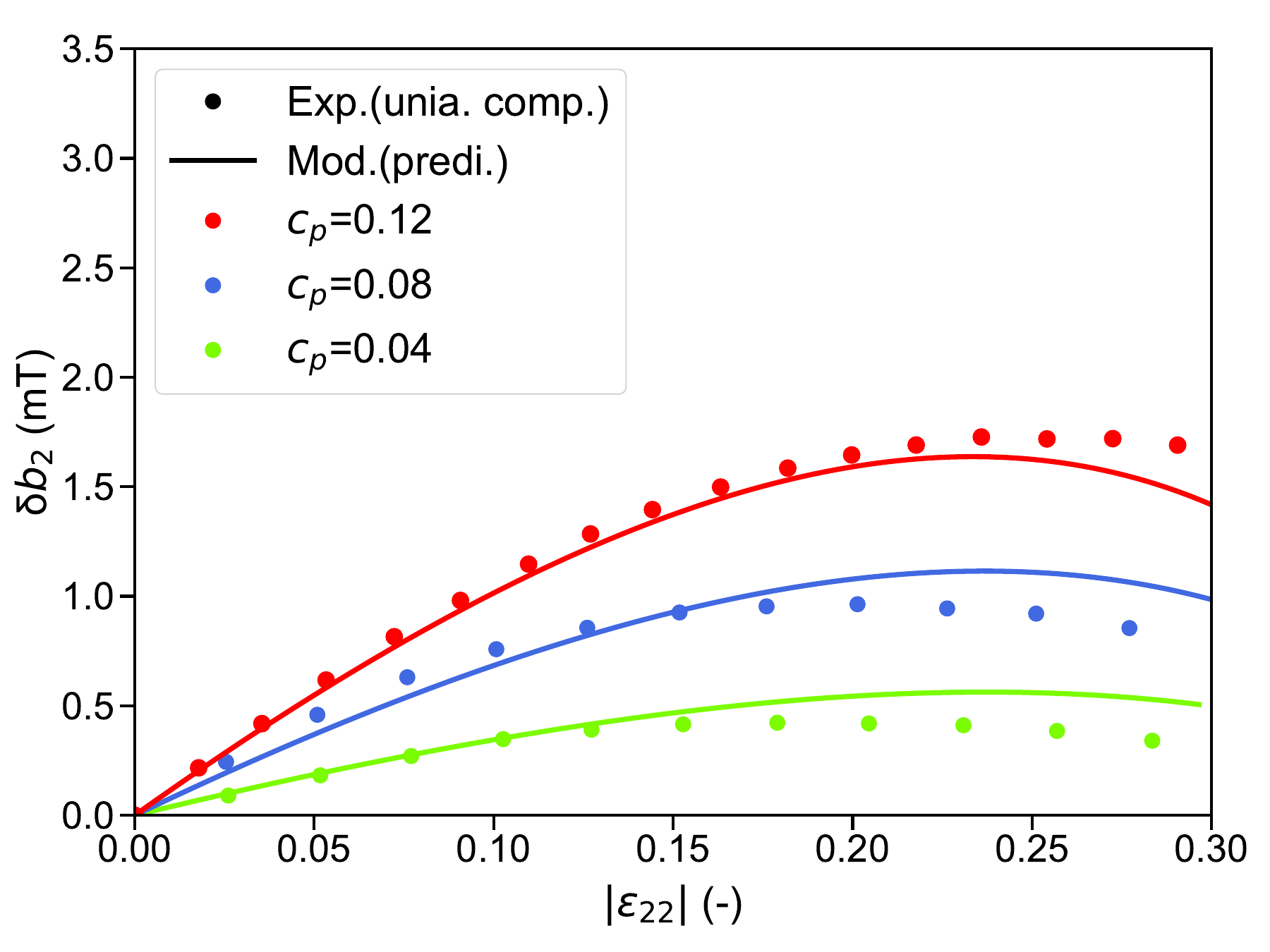}
  \end{subfigure}  
  \caption{Experiments (symbols) versus model (dashed for calibrated and continuous for prediction lines) for the magnetized samples subjected to oedometric compression and uniaxial compression with lateral traction-free boundary conditions. Current magnetic flux $b_2$ measured at the mid-bottom plane of the cube at an offset distance of $-0.5~\si{\milli \meter}$. (a) Average engineering compressive stress obtained as the measured force divided by the initial cube cross-section for both oedometric and uniaxial compression, (b) magnetic flux at $|\varepsilon_{22}|=0$ as a function of the particle volume fraction. The value corresponding at $c_\pt=0.12$ is used to calibrate $\mu_0 m^s_\pt=0.765~\si{\tesla}$ and the rest are predicted. Current magnetic flux variation along the magnetization direction for the (c) oedometric and (d) uniaxial compression tests. The response for $c_\pt=0.12$ (repeated from Fig.~\ref{fig:exp-mod_calibration_oedometric}) is used to calibrate the values $s_1$ and $s_2$ in \eqref{eq:Wmagr_def}, whereas the remaining curves in (c) and (d) are model predictions.}
  \label{fig:exp-mod_prediction_clamped}
\end{figure}

In this regard, the predictive capabilities of the model are assessed further by simulating samples with three particle volume fractions $c_\pt=0.04, 0.08, 0.12$ (see Table~\ref{tab:microstructural_params} for the complete set of microstructural parameters) under both the oedometric and the uniaxial compression loading conditions (see Fig.~\ref{fig:exp_setup}b). The corresponding comparisons are reported in Fig.~\ref{fig:exp-mod_prediction_clamped}. For the solution of the uniaxial compression experiment with lateral traction-free surfaces, we assume for simplicity that the top and bottom faces of the cube are subjected to pure Dirichlet boundary conditions (clamped), with a uniform compressive displacement imposed on the top surface. In addition, following the approach described in detail in \cite{psarra2019} and \cite{rambausek2022}, the surrounding air nodes are forced to deform following the deformation of the boundary of the solid. This imposed deformation regresses with the distance from the solid and becomes zero far enough from it. 

The corresponding numerical FE model results are compared with the corresponding experimental data in Fig.~\ref{fig:exp-mod_prediction_clamped}. The agreement between the model and experiments is extremely good for both the stress $|S_{22}|$ and magnetic field variation $\db_2$ and for all particle volume concentrations considered here without the need of an additional parameter or further calibration other than that carried out for $c_\pt=0.12$ in Fig.~\ref{fig:exp-mod_calibration_oedometric}. 

\begin{figure}[h!]
  \centering
   \includegraphics[width=0.9\textwidth, valign=t]{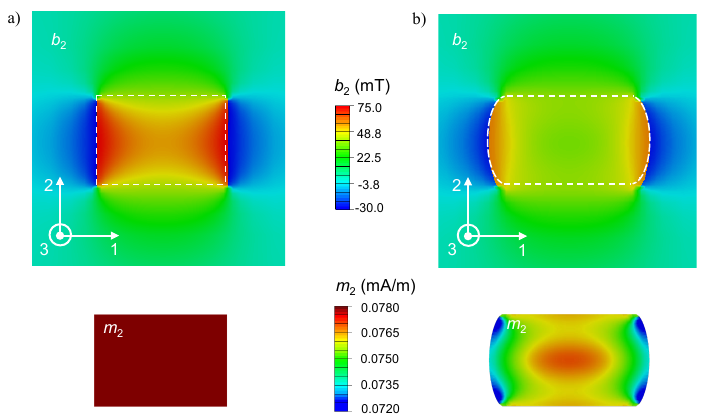}
  \caption{Magnetic flux $b_2$ and magnetization $m_2$ mid-plane FE contours of samples with $c_\pt=0.12$ for (a) the oedometric compression and (b) the uniaxial compression with traction-free lateral faces. The contours show the sample and a part of the surrounding air and correspond to a global applied compressive strain $\varepsilon_{22} = -0.3$. The dashed lines serve to indicate schematically the boundary of the $h$-MRE foam.}
  \label{fig:contours_oedometric_clamped}
\end{figure}

We remark at this point, that contrary to the oedometric experiment, in the uniaxial compression experiment, the resulting deformation of the foam is highly nonuniform (see contours in Fig.~\ref{fig:contours_oedometric_clamped}). In the actual uniaxial compression experiment, we observe some sliding of the upper and lower surfaces thus explaining the slightly softer response in the experimental results. The model estimates for $\db_2$ deviates weakly from the experimental curves for $\left| \varepsilon_{22} \right| > 0.2$, whereas the simulation stops for larger strains. This is directly attributed to the observation that, at larger compressive strains, the bulging of the lateral sides of the sample is too strong, thereby crushing the air elements near the top and bottom corners. This is a consequence of the limitations of our numerical approach, which currently cannot consider more realistic contact between the sample and the upper and lower rigid plates used to impose the deformation. Such extensions of the present numerical approach are beyond the scope of this study but are underway and hopefully will be presented elsewhere in the near future.

For completeness, we show in Fig.~\ref{fig:contours_oedometric_clamped} the mid-plane cross-sections of the deformed $b_2$ and $m_2$ contour fields of samples with $c_\pt=0.12$ obtained for (a) the oedometric compression and (b) the uniaxial compression with stress-free lateral sides BVPs. It is clear from those contours that the magnetic flux field is fairly complex and nonuniform in both cases, whereas the magnetization remains uniform even after large deformations in the oedometric test only. In turn, the modification of the \cite{Gou2004} solutions to take into account the deformation of the deformed cuboid in the case of the oedometric test allows for an elegant and analytical treatment of the BVP (including the solid and the surrounding air) which makes the calibration of the model easily attainable. 

\begin{remark}
The oedometric test is not appropriate for incompressible materials, such as the more common $h$-MREs, while the uniaxial compression test with lateral traction-free sides tends to deliver also nonuniform mechanical fields due to bulging. In that case, there exists no simple analytical solution neither for the magnetic fields nor the mechanical ones and thus a full three-dimensional FE analysis would have been required to calibrate any model. Nevertheless, a full three-dimensional solution is extremely time-consuming and thus it is not currently possible to use it in an optimizer setting to calibrate material parameters. Alternatively, for more common $h$-MREs (but even for $h$-MRE foams), one can resort to the more recent theoretical approach proposed by \cite{Stewart2025} or to an entirely different test that allows to obtain analytically model parameters as presented in the recent work of \cite{Danas2025} and corresponds to a simple shear experiment with no air gap between the magnetic poles and the specimen. 
\end{remark}

\begin{figure}[h!]
  \centering
  \begin{subfigure}[t]{0.02\textwidth}
    {\small a)}\hfill
  \end{subfigure}
  \begin{subfigure}[t]{0.45\textwidth}
    \includegraphics[width=\linewidth, valign=t]{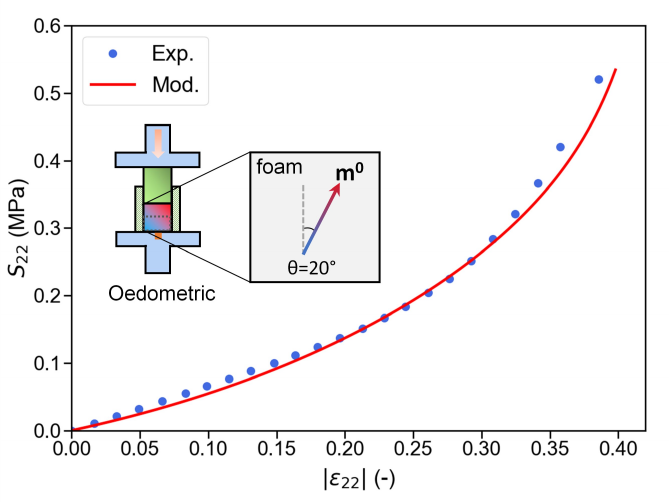}
  \end{subfigure}\hfill
  \begin{subfigure}[t]{0.02\textwidth}
    {\small b)}\hfill
  \end{subfigure}
  \begin{subfigure}[t]{0.45\textwidth}
    \includegraphics[width=\linewidth, valign=t]{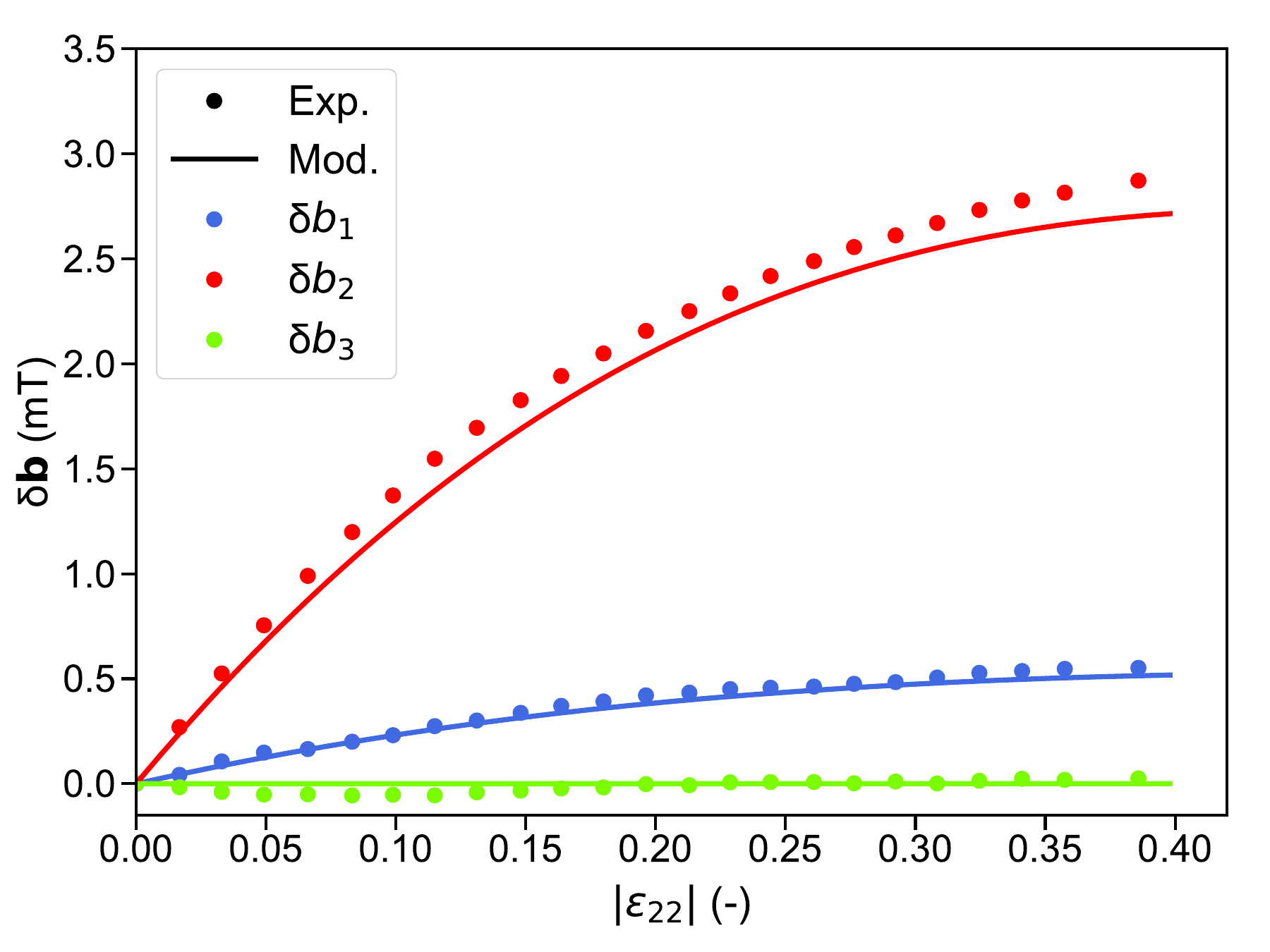}
  \end{subfigure}
  \caption{Experiments (symbols) versus model (cont. lines) predictions for the magnetized sample with $c_\pt=0.12$ pre-magnetized at an angle $\theta=20^o$ with respect to the 2-axis as shown in the inset and subjected to oedometric boundary conditions. Current magnetic flux $b_2$ measured at the mid-bottom plane of the cube at an offset distance of $-0.5~\si{\milli \meter}$. (a) Average engineering compressive stress for the oedometric loading conditions, and (b) current magnetic flux variations of all three components of $\bvb$.}
  \label{fig:exp-mod_angle_theta}
\end{figure}
For further assessment, we close this section with Fig.~\ref{fig:exp-mod_angle_theta}, which shows oedometric model predictions versus experimental data for pre-magnetized samples at an angle $\theta=20^o$ with respect to the 2-axis (as shown in the inset sketch). For the pre-magnetization, we have used an impulse magnetizer (DXMM-12C40 Magnetizer, Dexing Magnet Tech. Co., Limited, China) capable of delivering a field above $5~\si{\tesla}$ at the center of a $50~\si{\milli \meter}$ coil in 2~s (see \cite{Perez-Garcia_2024}). Due to the misalignment of the magnetization with respect to the main cubic axes and compression direction, one obtains magnetic flux variation also along direction 1, thus leading to a non-zero $\db_1$ in addition to $\db_2$ ($\db_3=0$ since no pre-magnetization exists along this direction). Note that in this case, we cannot use the analytical \cite{Gou2004} expressions as they require that the magnetization is aligned with the cubic axes. We observe that the model is capable of predicting this non-trivial case too with very good accuracy for both the stress and the magnetic flux variation. With compression, $\db_1$ increases and saturates similar to $\db_2$. It is worth noticing here that while the \emph{normal} component $b_2$ is continuous at the bottom face of the cube, the tangential component $b_1$ (and hence $\db_1$) exhibits a jump as one moves from the cube towards the air. Given the rigorous modeling and experimental analysis of both the solid and the surrounding air in the present approach, this result is properly captured both by the model and the sensor used to measure the fluxes in the experiments.

\section{Numerical study: multi-loading sensing capabilities of $h$-MRE foam}
\label{sec:Numerical study: multi-loading sensing capabilities of hMRE foam}

In this section, we use the FE model to explore the $h$-MRE foam response under more complex compression/tension plus shear loads as well as nonuniform compressive loads. For brevity, we focus on one particle volume fraction, $c_\pt=0.12$ (and $c_\vt=0.45$ and $c_\pmt=0.21$).

The goal is to show  that the three-dimensional magnetic flux variations allow to infer directly the input deformation mode, i.e., to distinguish between tension and compression as well as shear loads in different directions. 

In this view, we consider first a uniformly applied tension/compression with shear load at the top surface of the sample, such that the overall (effective) engineering strains are given by
\begin{equation}
	\varepsilon_{22}=\pm 0.1t, \quad \gamma_{12}=0.2t, \quad \gamma_{32}=0.1t, \qquad t\in[0,1].
\end{equation}
The positive sign in $\varepsilon_{22}$ leads to tension and the negative sign to compression. 
\begin{figure}[h!]
  \centering
  \begin{subfigure}[t]{0.02\textwidth}
    {\small a)}
  \end{subfigure}
  \begin{subfigure}[t]{0.3\textwidth}
    \includegraphics[width=\linewidth, valign=t]{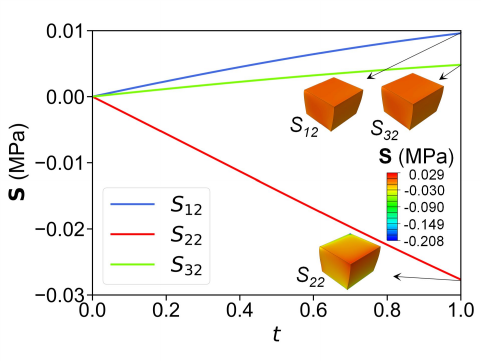}
  \end{subfigure}\hfill
  \begin{subfigure}[t]{0.02\textwidth}
    {\small b)}
  \end{subfigure}
  \begin{subfigure}[t]{0.3\textwidth}
    \includegraphics[width=\linewidth, valign=t]{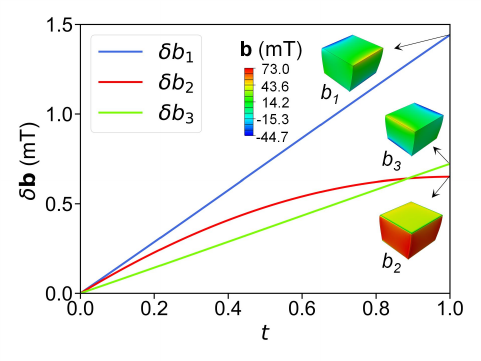}
  \end{subfigure}\hfill
  \begin{subfigure}[t]{0.02\textwidth}
    {\small c)}
  \end{subfigure}
  \begin{subfigure}[t]{0.3\textwidth}
    \includegraphics[width=\linewidth, valign=t]{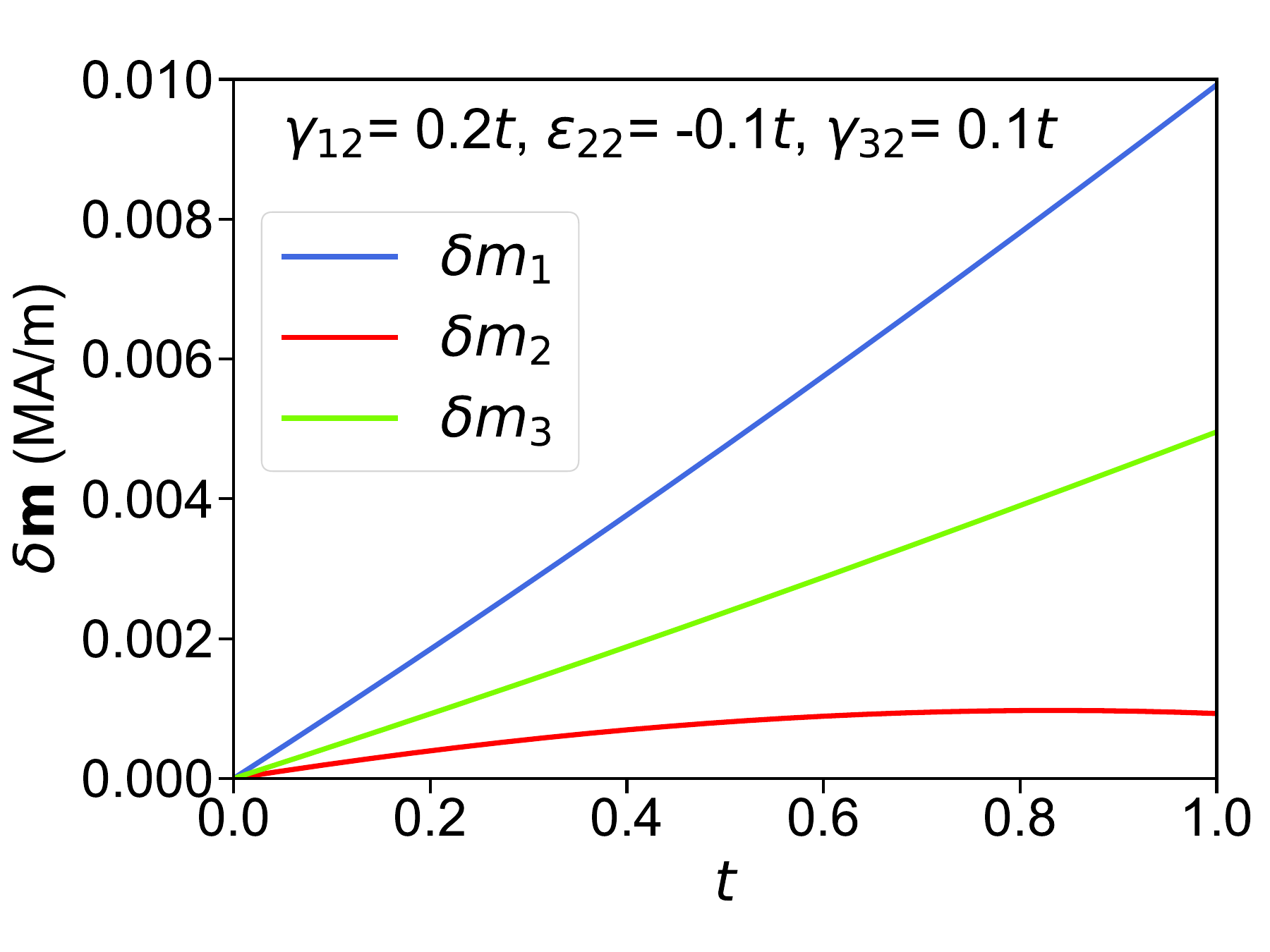}
  \end{subfigure}
  \\[2ex]
  \begin{subfigure}[t]{0.02\textwidth}
    {\small d)}
  \end{subfigure}
  \begin{subfigure}[t]{0.3\textwidth}
    \includegraphics[width=\linewidth, valign=t]{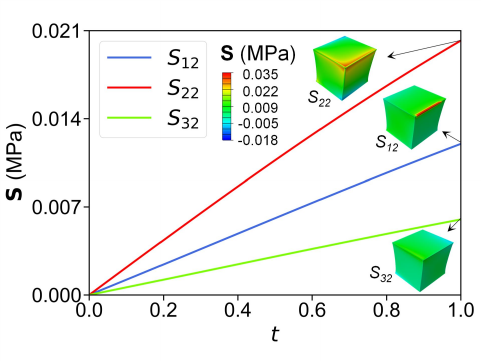}
  \end{subfigure}\hfill
  \begin{subfigure}[t]{0.02\textwidth}
    {\small e)}
  \end{subfigure}
  \begin{subfigure}[t]{0.3\textwidth}
    \includegraphics[width=\linewidth, valign=t]{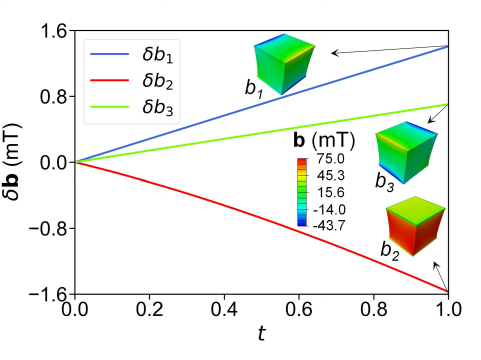}
  \end{subfigure}\hfill
  \begin{subfigure}[t]{0.02\textwidth}
    {\small f)}
  \end{subfigure}
  \begin{subfigure}[t]{0.3\textwidth}
    \includegraphics[width=\linewidth, valign=t]{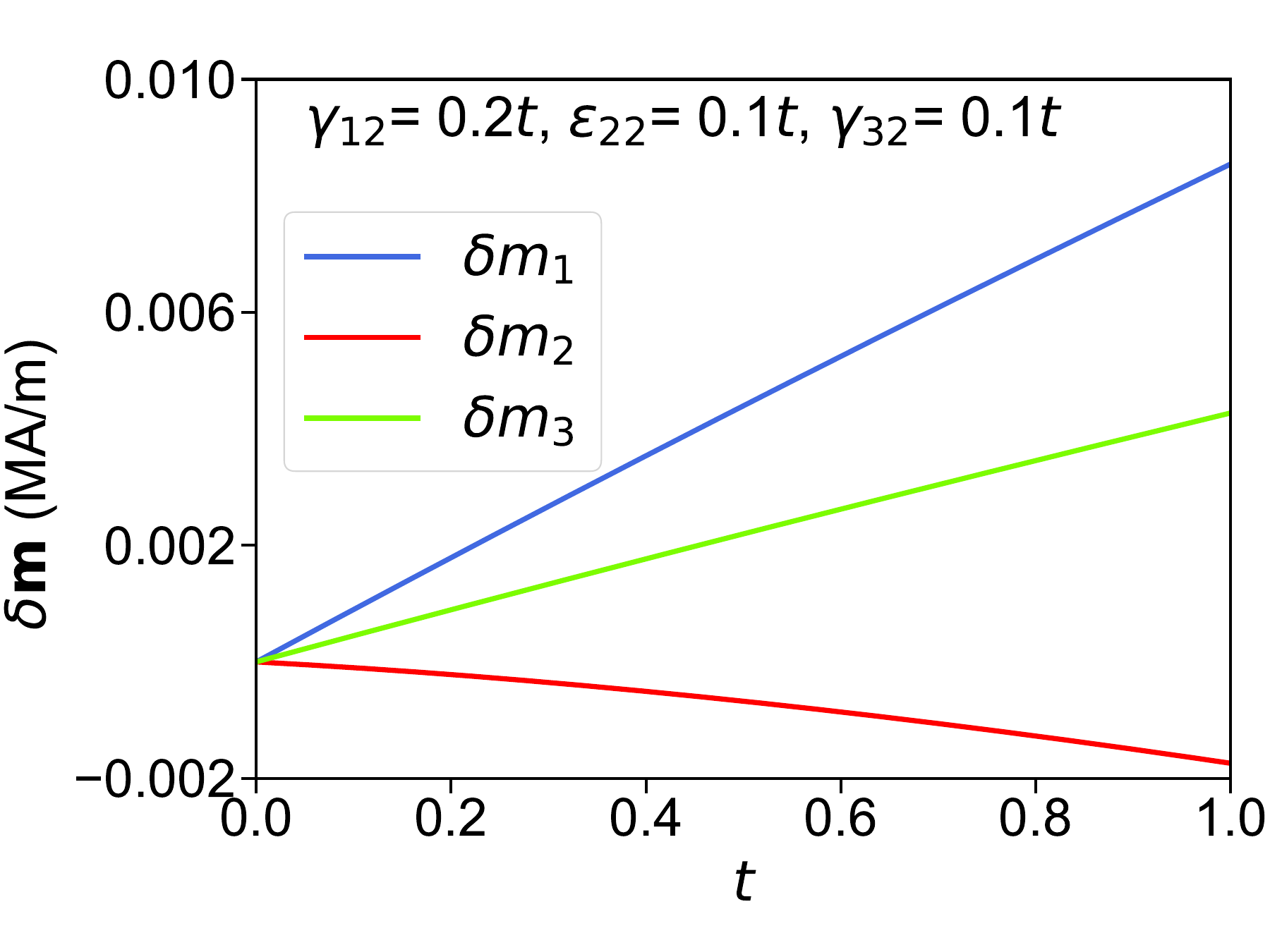}
  \end{subfigure}\hfill  
  \caption{Numerical simulations for combined compression (a--c) and tension (d--f) along $2$ and simple shear loading $12$ and $32$. (a) Average over the cube first Piola stress components, (b) current magnetic flux variation $\delta \bvb$ (all three components change due to shearing) at the mid-bottom of the sample and (c) average magnetization variation $\delta \mvb$ over the sample as a function of the applied loading amplitude $t \in [0,1]$ with (a--c) $\gamma_{12}=0.2t$, $\varepsilon_{22}=-0.1t$ and $\gamma_{32}=0.1t$ and (d--f) $\gamma_{12}=0.2t$, $\varepsilon_{22}=0.1t$ and $\gamma_{32}=0.1t$. Corresponding contours indicating the nonuniform stress, magnetic flux and magnetization fields during the loading are shown as insets.}
  \label{fig:FE_compression_tension_shear}
\end{figure}
Figures~\ref{fig:FE_compression_tension_shear}a,d show the average engineering ($S_{22}$) and shear ($S_{12}$ and $S_{32}$) stress components, (b,e) the magnetic flux variation components $\db_1$, $\db_2$ and $\db_3$ and (c,f) the magnetization variation components $\dm_1$, $\dm_2$ and $\dm_3$ for (a--c) compression and (d--f) tension combined with the same state of shear. Given the different shear amplitudes, the resulting stress response is entirely expected. More interestingly, however, are the corresponding $\delta\bvb$ measures, which allow to distinguish between the different applied mechanical loads not only qualitatively but also quantitatively. For instance, a compression load along direction $2$ leads to a positive variation of $\db_2$, while $\db_1>\db_3$ as a result of $\gamma_{12}>\gamma_{32}$ (note that the first component of the shear strain corresponds to the direction of the shear load). In other words, one is able to identify what type of load and at what relative magnitude it is applied (on average) on the sample just by extracting the local magnetic flux at its mid-bottom plane. Similarly, if one applies tension instead of compression (see Fig.~\ref{fig:FE_compression_tension_shear}e red line and compare with ~\ref{fig:FE_compression_tension_shear}b red line), $\db_2<0$ implying that the voids are growing and the specimen is extending. Very similar conclusions may be drawn from the corresponding average magnetization response (Fig.~\ref{fig:FE_compression_tension_shear}c,f). The magnetization, however, is not immediately available at the boundary and hence is not a direct experimental evidence of the imposed deformation and in general requires further measurements to be obtained.

\begin{figure}[h!]
  \centering
  \begin{subfigure}[t]{0.02\textwidth}
    {\small a)}\hfill
  \end{subfigure}
  \begin{subfigure}[t]{0.45\textwidth}
    \includegraphics[width=\linewidth, valign=t]{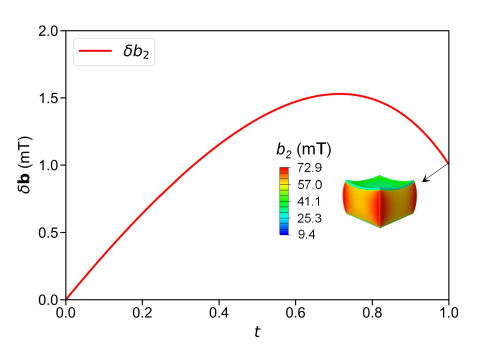}
  \end{subfigure}\hfill
  \begin{subfigure}[t]{0.02\textwidth}
    {\small b)}\hfill
  \end{subfigure}
  \begin{subfigure}[t]{0.45\textwidth}
    \includegraphics[width=\linewidth, valign=t]{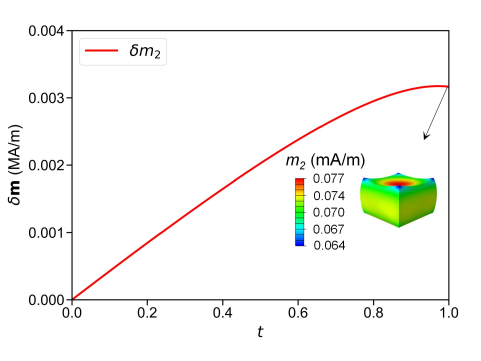}
  \end{subfigure}\\[2ex]
  \begin{subfigure}[t]{0.02\textwidth}
    {\small c)}\hfill
  \end{subfigure}
  \begin{subfigure}[t]{0.45\textwidth}
    \includegraphics[width=\linewidth, valign=t]{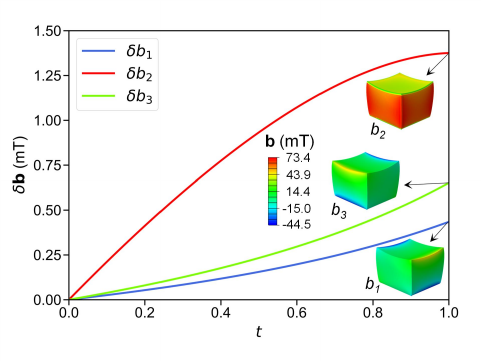}
  \end{subfigure}\hfill
  \begin{subfigure}[t]{0.02\textwidth}
    {\small d)}\hfill
  \end{subfigure}
  \begin{subfigure}[t]{0.45\textwidth}
    \includegraphics[width=\linewidth, valign=t]{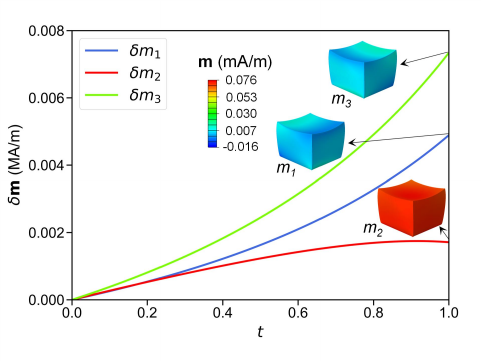}
  \end{subfigure}\hfill
  \caption{Numerical simulations for nonuniform quadratic compression applied (a,b) symmetrically and (c,d) asymmetrically along direction $2$. The exact displacement profile is defined in equation \eqref{eq:disp_quad_profile_paraboloid}. (a,c) Current magnetic flux variation $\delta \bvb$ at the mid-bottom of the sample (offset by $-0.5$~mm) and (b,d) average magnetization variation $\delta \mvb$ over the sample as a function of the applied loading amplitude $t \in [0,1]$. Corresponding contours indicating the magnetic flux and magnetization fields during the loading are shown as insets.}
  \label{fig:paraboloid_loading}
\end{figure}

As a second example, we consider a nonuniform compressive load with a quadratic imprint on the top surface of the cubic sample serving to mimic the compression of a stiff finger-type piston. The applied normal component of the displacement, $u_2$, is assumed to be a function of the reference in-plane coordinates $X_1$ and $X_3$ such that 
\begin{equation}
u_2^{\mathtt{top}}(X_1,X_3) = \frac{(X_1 - A_1-X_1^\mathtt{mid})^2 + (X_3 - A_3-X_3^\mathtt{mid})^2}{4 R} + u_2^\mathtt{app} t, \qquad t\in[0,1].
\label{eq:disp_quad_profile_paraboloid}
\end{equation}
In this equation, $X_1^\mathtt{mid}$ and $X_3^\mathtt{mid}$ denote the reference coordinates of the center of mass of the cube, $R=5$~mm is a reference radius controlling the extent of the quadratic function and $t$ is a non-dimensional loading parameter. The tangent directions on the top surface are assumed traction-free allowing to reach larger strains. We consider next two cases, a fully symmetric and an asymmetric one. In the first case, we set $A_1=A_3=0$ and $u_2^\mathtt{app}=-3.5~$mm thus inducing a fully symmetric quadratic displacement profile at the top surface. In the second case, we choose $A_1=2$~mm, $A_3=3$~mm and $u_2^\mathtt{app}=-2.5$~mm, which leads to an offset of the top mid-plane displacement profile inducing this way non-zero and nonuniform shear strains. In Fig.~\ref{fig:paraboloid_loading}, we show the resulting current magnetic flux variation $\delta \bvb$ at the mid-bottom surface of the cube (at an offset of $-0.5$~mm) and the average magnetization variation in the sample for the (a,b) symmetric and the (c,d) asymmetric cases, respectively.

Similar to the previous uniform tension/compression case, we observe that the symmetric case in Fig.~\ref{fig:paraboloid_loading}a leads to zero tangent magnetic flux variations $\db_1=\db_3=0$, whereas the asymmetric compression in Fig.~\ref{fig:paraboloid_loading}c induces nonuniform shearing and thus leads to non-zero $\db_3>\db_1>0$ response. In particular, it allows to infer that the deformation profile is biased more towards direction $3$ than direction $1$. Again, due to compressive loads, $\db_2>0$ is positive and non-monotonic similar to the uniaxial compression experiment discussed in Fig.~\ref{fig:exp-mod_prediction_clamped}. It is very interesting to observe at this point that, due to the spatial nonuniformity of the applied load, the resulting increase of $\db_1$ and $\db_3$ is much more nonlinear than the corresponding shear responses in Figs.~\ref{fig:FE_compression_tension_shear}b,e. This implies that qualitatively different loading conditions appear directly and quantitatively on the measured magnetic fluxes. The corresponding magnetization variations overall exhibit similar trends with the magnetic fluxes. We note however that the normal component $\dm_2$ is lower in amplitude than the tangential components.

\section{Concluding remarks}
\label{sec:Concluding remarks}

The present study has shown for the first time experimentally and theoretically the potential of using bulk $h$-MRE foams to produce measurable magnetic fluxes in the surrounding space when deformed. We have first fabricated fairly isotropic $h$-MRE foams with various NdFeB particle volume fractions and resulting porosities. Those foams have been permanently magnetized in sufficiently large magnetic fields larger than $3~\si{\tesla}$ to reach full magnetization saturation. Subsequently, we have carried out two compression experiments; the first one is known as the oedometric test and leads to uniform uniaxial strain and magnetization fields (but not magnetic fluxes and strengths) thus allowing for an analytical treatment of the magneto-mechanical problem (provided that the magnetization is aligned with the cubic axes) even when the magnetic fields are not uniform as is the case here. The second experiment instead is used as a control case for validation of the proposed model and is a standard uniaxial compression test with traction-free lateral surfaces. Yet, due to the highly adhesive nature of the contact between the compressive device and the sample, this experimental setup leads to highly nonuniform and non-analytical mechanical and magnetic fields. Those tests have revealed unambiguously the potential of $h$-MRE foams in providing measurable magnetic flux variations (in the order of $1{-}4~\si{\milli \tesla}$) in their immediate surrounding when deformed. 

Subsequently, we have proposed a novel compressible magneto-elastic foam model with magnetic dissipation using a combination of known mechanical models for porous polymers and earlier constitutive descriptions for incompressible $h$-MREs. The proposed foam model is based on homogenization results and has only two additional parameters, which however have been shown to be sufficiently calibrated only with one particle volume fraction case. Specifically, most of the resulting coefficients involve directly the reference particle NdFeB volume fraction, $c_\pt$, and the particles' magnetic properties (such as coercivity, magnetization saturation, etc) but further testing is needed especially for the dissipative part, which has been of less relevance in this work and has not been pursued further (see some recent work by \cite{Lin2025} along this direction). In turn, one of the magneto-mechanical terms involves directly coupling with the third invariant of the deformation tensor, i.e., $J=\det\Fb$. This term allows to predict the apparent magnetization changes that are directly related to pore closure or opening. The first leads to reduction of $J<1$ and increase of the total particle volume fraction $c_\pt$ and thus increase of the effective saturation magnetization and resulting magnetic flux. In turn, pore opening implies $J>1$ and thus decrease of $c_\pt$ and the corresponding saturation magnetization. Those mechanisms, inherent in such magnetic foams, are not present in the more common incompressible $h$-MREs. 

The proposed model has then been coupled with the \cite{Gou2004} solutions for cuboid permanent magnets (which had to be extended to include finite strains) and the oedometric data to identify two coefficients related to the presence of $J$ and the particle saturation magnetization. This has provided a robust model that is able to recover both the stress and magnetic flux distributions in the $h$-MRE foam during the oedometric test. Subsequently, the model has been numerically implemented in three dimensions and used to probe additional oedometric and uniaxial compression experiments for different particle volume fractions showing sufficient accuracy without the need of recalibration. 

The proposed model has then been used to study a selected set of cases promoting the versatility of responses and its inherent magneto-mechanical coupling. In particular, we have shown for the first time that this material can be used as a potential multi-modal soft tactile/haptic sensor. We have revealed by means of numerical simulations that when the sample is compressed/stretched and sheared with different amplitudes, this results in clearly distinguishable magnetic flux measurements in the surroundings. Thus, by measuring these magnetic fluxes, which is fairly easy and can be integrated in any device of interest, one is able to infer whether the sample has been uniformly or nonuniformly compressed or stretched, sheared and in which direction. All these deformation modes are inherently encompassed in a single and geometrically simple cubic $h$-MRE foam sample. 

Finally, the theoretical analysis combined with the experimental part of this work has revealed a nontrivial dependence of the current magnetization on the Jacobian of the deformation gradient and one more time the fact that magnetization is a constitutive variable that has to be identified according to the given material at hand. Moreover, it is straightforward to show that a simplistic transformation-type $\mvb=J^{-1}\Fb \mvb^0$ law (with $\mvb^0$ denoting the pre-magnetization vector whose magnitude is equal to $|\mvb^0|=m^s$ in the present case) would have been entirely insufficient for such highly compressible solids. The reason is simply that in the case of the oedometric test, it would have led to zero magnetization changes, which is contrary to the current experiments and even intuition. 

This study opens the way for new multi-modal haptic sensing capabilities by use of mechanically-soft $h$-MRE foams. The resulting magnetic fluxes are easily measurable in the surrounding of the sample allowing for a direct way of obtaining information about finite mechanical deformation of versatile type, e.g., combined compression with shears at different directions and different amplitudes as well as distinction between uniform and nonuniform loads.

\section*{Acknowledgements}
Zehui Lin acknowledges the China Scholarship Council for his doctoral funding. All authors acknowledge support from the European Research Council (ERC) under the European Union's Horizon 2020 research and innovation program (grant agreement No 101081821). The authors thank  Dr. Lopez-Donaire and Prof. Garcia-Gonzalez for their help in magnetizing the \textit{h}-MRE foam samples at an angle $\theta=20^o$ with their magnetizer at U. Carlos III, Madrid allowing to complete the data in Fig.~\ref{fig:exp-mod_angle_theta}.

\appendix

\section{Analytical solution of Gou et al.}
\label{Appx:Analytical solution of Gou et al.}

The remaining components from \cite{Gou2004} solution lying at the plane perpendicular to the applied magnetization read
\begin{align}
b_1^r(x_1,x_2,x_3)=\dfrac{\mu_0 \,m_2}{8\pi}\Big[
   &F_{13}\left(x_1,x_2,L_3-x_3\right)+F_{13}\left(x_1,L_2-x_2,x_3\right)\nonumber\\[1ex]
   -&F_{13}\left(x_1,L_2-x_2,L_3-x_3\right)-F_{13}\left(L_1-x_1,x_2,x_3\right)\nonumber\\[1ex]
   +&F_{13}\left(L_1-x_1,x_2,L_3-x_3\right)+F_{13}\left(L_1-x_1,L_2-x_2,x_3\right)\nonumber\\[1ex]
   -&F_{13}\left(L_1-x_1,L_2-x_2,L_3-x_3\right)-F_{13}(x_1,x_2,x_3)\Big]
\end{align}
\begin{align}
b_3^r(x_1,x_2,x_3)=\dfrac{\mu_0 \,m_2}{8\pi}\Big[
   &F_{13}\left(x_3,x_2,L_1-x_1\right)+F_{13}\left(x_3,L_2-x_2,x_1\right)\nonumber\\[1ex]
   -&F_{13}\left(x_3,L_2-x_2,L_1-x_1\right)-F_{13}\left(L_3-x_3,x_2,x_1\right)\nonumber\\[1ex]
   +&F_{13}\left(L_3-x_3,x_2,L_1-x_1\right)+F_{13}\left(L_3-x_3,L_2-x_2,x_1\right)\nonumber\\[1ex]
   -&F_{13}\left(L_3-x_3,L_2-x_2,L_1-x_1\right)-F_{13}(x_3,x_2,x_1)\Big]
\end{align}
with $m_2$ given by \eqref{eq:m2_oedometric_sol} and $F_{13}$ being a non-dimensional function of the spatial coordinates given by
\begin{align}
F_{13}(f_1,f_2,f_3)=
\begin{cases}
\log\left[\dfrac{\sqrt{f_1^2+f_2^2+f_3^2}-f_1}{\sqrt{f_1^2+f_2^2+f_3^2}+f_1}\right], \qquad f_1, f_2, f_3 \neq 0\\[2ex]
0, \qquad \text{otherwise.}
\end{cases}
\end{align}

\bibliographystyle{elsarticle-harv}

\end{document}